\documentclass[pre,twocolumn,amsmath,amssymb,nofootinbib,floatfix,superscriptaddress]{revtex4}

\usepackage{graphicx,bm,xcolor, bbold}
\usepackage{enumerate}
\usepackage{graphicx,bm}
\usepackage[normalem]{ulem}
\usepackage{float}

\makeatletter
\def\graphicscale{\twocolumn@sw{0.3}{0.4}}
\def\graphicthreescale{\twocolumn@sw{0.3}{0.4}}

\begin{document}

\title{ Out-of-equilibrium spinodal-like scaling behaviors across
  \\ the magnetic first-order transitions of 2D and 3D Ising systems}

\author{Andrea Pelissetto}
\affiliation{Dipartimento di Fisica dell'Universit\`a di Roma Sapienza
        and INFN Sezione di Roma I, I-00185 Roma, Italy}

\author{Ettore Vicari} 
\affiliation{Dipartimento di Fisica dell'Universit\`a di Pisa,
        Largo Pontecorvo 3, I-56127 Pisa, Italy}

\date{\today}

\begin{abstract}
  We study the out-of-equilibrium scaling behavior of two-dimensional
  and three-dimensional Ising systems, when they are slowly driven
  across their {\em magnetic} first-order transitions at low
  temperature $T<T_c$, where $T_c$ is the temperature of
  their continuous transition.  We consider Kibble-Zurek (KZ)
  protocols in which a spatially homogenous magnetic field $h$ varies
  as $h(t)=t/t_s$ with a time scale $t_s$. The KZ dynamics starts from
  negatively-magnetized configurations equilibrated at $h_i<0$ and
  stops at a positive value of $h$ where the configurations acquire a
  positive average magnetization. We consider the Metropolis and the
  heat-bath dynamics, which are two specific examples of a purely
  relaxational dynamics.  We focus on two different dynamic
  regimes. We consider the out-equilibrium finite-size scaling (OFSS)
  limit in which the system size $L$ and the time scale $t_s$ diverge
  simultaneously, keeping an appropriate combination fixed. Then, we
  analyze the KZ dynamics in the thermodynamic limit (TL), obtained by
  taking first the $L\to\infty$ limit at fixed $t$ and $t_s$, and then
  considering the scaling behavior in the large-$t_s$ limit. Our
  numerical results provide evidence of OFSS, as predicted by general
  scaling arguments.  The results in the TL show the emergence of
  spinodal-like behaviors: The passage from the negatively-magnetized
  phase to the positively-magnetized one occurs at positive values
  $h_*>0$ of the magnetic field, which decrease as $h_* \sim 1/(\ln
  t_s)^\kappa$, with $\kappa = 2$ and $\kappa=1$ in two and three
  dimensions, respectively, for $t_s\to\infty$. We identify $\sigma
  \equiv t (\ln t)^\kappa/t_s$ as the relevant scaling variable
  associated with the KZ dynamics in the TL.
  \end{abstract}

\maketitle

\section{Introduction}
\label{intro}

Many-body systems driven across phase transitions show an
out-of-equilibrium behavior, even when the system parameters are
varied very slowly, because large-scale modes are not able to
equilibrate on any time scale.  Hysteresis and coarsening phenomena,
critical aging, Kibble-Zurek (KZ) defect production are typical
examples of out-of-equilibrium phenomena that are observed at
continuous and discontinuous transitions in many different, classical
and quantum
contexts~\cite{Kibble-76,Kibble-80,Binder-87,Bray-94,Zurek-96,CA-99,
  CG-04,PSSV-11,CEGS-12,
  Biroli-16,RV-21,PV-24,Zurek-85,CDTY-91,BCSS-94,BBFGP-96,Ruutu-etal-96,
  CPK-00,CGMB-01,MMR-02,
  MPK-03,ZDZ-05,CGM-06,MMARK-06,SHLVS-06,PG-08,WNSBDA-08,GPK-10,CWCD-11,
  Chae-etal-12,MBMG-13,EH-13,Ulm-etal-13,Pyka-etal-13,LDSDF-13,
  Corman-etal-14,NGSH-15,Braun-etal-15,DWGGP-16,RV-20}.

A notable example is the KZ dynamics~\cite{Kibble-80,Zurek-96}, in
which a system parameter $h$ (such as an external {\em magnetic} field
or the reduced temperature at thermal phase transitions) varies as
$h(t)=t/t_s$ across a transition point $h=h_c=0$, where $t_s$ is a
time scale. At continuous transitions the evolution of the system
under the KZ dynamics develops an out-of-equilibrium scaling behavior
for large values of $t_s$.  The corresponding critical power laws are
related with the length-scale critical exponent $\nu$, which only
depends on the static universality class of the transition, and with
the dynamic exponent $z$, which also depends on the type of
dynamics. The KZ dynamics has been studied both in the thermodynamic
limit (TL)~\cite{Kibble-76,Kibble-80,Zurek-85,Zurek-96,ZDZ-05,PG-08,
  PSSV-11,CEGS-12} and in finite systems~\cite{RV-21,TV-22,DV-23}.
Finite-size systems develop an out-of-equilibrium finite-size scaling
behavior (OFSS), which can be related to the out-of-equilibrium
scaling in the TL by a straightforward infinite-size
limit~\cite{RV-21}, due to the fact that the critical modes that
control the scaling behaviors are the same in finite systems and in
the~TL.

KZ or more general quenching protocols have been also studied at
first-order classical and quantum
transitions~\cite{Binder-87,Pfleiderer-05,PV-24}, showing more complex
out-of-equilibrium behaviors, see, e.g.,
Refs.~\cite{RV-21,PV-24,MM-00,BBD-08,LFGC-09,AC-09,YKS-10,JLSZ-10,NIW-11,
  TB-12,ICA-14,PV-15,PV-16,PV-17,PV-17b,LZ-17,PPV-18,SW-18,PRV-18,
  PRV-18-def,Fontana-19,LZW-19,PRV-20,DRV-20,CCP-21,SCD-21,
  CCEMP-22,TS-23,Surace-etal-24,PRV-25,PRV-25b}.  In particular,
qualitatively different mechanisms work in a finite volume and in the
TL, giving rise to unrelated scaling behaviors in the two cases. In
particular, an OFSS theory has been developed to provide a scaling
description of the dynamic evolution of finite-size systems in the
large-$t_s$ limit (it extends in a natural way the finite-size scaling
theory that has been
developed~\cite{NN-75,FB-82,PF-83,FP-85,CLB-86,BK-90,LK-91,BK-92,VRSB-93,
  CNPV-14,CNPV-15,CPV-15,CPV-15-iswb,PRV-18-fowb,YCDS-18,RV-18} to
describe the static equilibrium properties of systems close to the
transition).  The scaling behavior of the KZ dynamics in the TL is
instead much less understood and shows distinct peculiar scaling
features.

The scaling properties of the KZ dynamics in finite systems and in the
TL were discussed in Refs.~\cite{PV-17,PPV-18,PV-24} at the thermal
(classical) first-order transition occurring in the 2D $q$-state Potts
model for $q>4$~\cite{Baxter-book,Wu-82}. Simulations were performed
for $q=20$ and $q=10$, varying the temperature across the thermal
first-order transition starting from the disordered high-temperature
phase.  They show the emergence of a spinodal-like\footnote{In the
standard theory of first-order transitions one defines the spinodal
line as the line where the metastable state becomes unstable (the
free-energy minimum corresponding to the metastable state
disappears). In the mean-field approach the spinodal line is located
at a finite value of the external field $h$.  However, in realistic
short-ranged models there is no metastable state for any $h > 0$ in
the infinite-size limit~\cite{Binder-87}.  From a dynamic perspective,
in the limit of very slow adiabatic variations of the parameters in
large-size systems, it is not possibile to observe quasi-equilibrium
metastable states in short-ranged model when $h$ is finite.
Spinodal-like behaviors can instead br observed in an appropriate
out-of-equilibrium scaling regime~\cite{PV-17,PV-24}, as also
discussed in this paper. \label{footnote1}} scaling
behavior~\cite{PV-17} in the TL.  Therefore it is interesting to check
whether an analogous behavior occurs at other classical first-order
transitions, and, in particular, if it also occurs in 3D systems, to
deepen our understanding of the general features of the KZ dynamics at
classical first-order transitions.

These issues have also been recently investigated at first-order
quantum transitions.  For example, Refs.~\cite{PRV-25,PRV-25b}
analyzed the KZ dynamics in the one-dimensional quantum
Ising model in the presence of a transverse static field $g$ and of a
longitudinal time-dependent field $h$, driving the system across the
first-order transition line present for small values of
$g$~\cite{PRV-25,PRV-25b}.  The results show that the KZ evolution in
the TL is not related with the variety of OFSS behaviors observed in
finite systems, which depend on the boundary
conditions~\cite{RV-21,PV-24}.  Apparently, the mechanisms that are at
the basis of the OFSS behavior are not relevant in the
TL~\cite{PRV-25,PRV-25b}.  Therefore, at variance with what occurs at
continuous quantum transitions, the out-of-equilibrium behaviors that
occur in the TL are distinct from those observed in the OFSS limit,
see, e.g., Refs.~\cite{PRV-25,PRV-25b,SCD-21}.

In this paper we extend the previous studies to classical {\em
  magnetic} first-order transitions. We consider Ising models in the
presence of a spatially homogeneous magnetic field $h$, which drives
first-order transitions in the low-temperature phase, below the
critical temperature $T_c$ where the model undergoes a continuous
transition.  We study the out-of-equilibrium behavior arising from the
simplest KZ protocol, in which $h$ varies as $h(t)=t/t_s$, where $t_s$
is a time scale (the temperature $T<T_c$ is kept fixed). The KZ
dynamics starts at $t=t_i<0$ from an ensemble of configurations
equilibrated at $T<T_c$ and $h=h_i= h(t_i)<0$, so the initial
magnetization $m$ is negative. Then, the system evolves under a
relaxational dynamics (model-A dynamics in the classification of
Ref.~\cite{HH-77}) up to positive values of $h(t)$, where $m(t)$
becomes eventually positive.  Specifically, we consider the standard
Metropolis and heat-bath dynamics~\cite{Binder-76}, which are commonly
used in Monte Carlo (MC) simulations. The time evolution is monitored
by computing averages of observables, such as the magnetization and
the bond-energy density, as a function of time.

We focus on two different dynamic regimes. First, we study the OFSS
behavior occurring in finite systems with periodic boundary
conditions. In this case, the relevant time scale is the exponentially
large time needed to observe the tunneling of the system from one
phase to the other at $h=0$. Then, we consider the KZ dynamics in the
TL, obtained by first taking the infinite-size limit keeping $t$ and
$t_s$, and thus $h(t)$, fixed, and then the large-$t_s$ limit.

Our numerical results confirm that the KZ dynamics develops an OFSS
behavior in finite systems. The time scale is the time needed for the
finite-size system to change phase in the absence of an external
magnetic field.  Our numerical analyses in the TL show the emergence
of a spinodal-like behavior: The magnetization changes sign at a
positive value $h=h_*>0$, which decreases as $t_s$ increases.  More
precisely, we find $h_* \sim 1/(\ln t_s)^\kappa$ with $\kappa = 2$ for
two-dimensional (2D) Ising systems, $\kappa=1$ for three-dimensional
(3D) Ising systems, and $\kappa=1/2$ for four-dimensional (4D) Ising
systems.

The paper is organized as follows.  In Sec.~\ref{moddyn} we present
the lattice models and the KZ protocol that we consider.  In
Sec.~\ref{outsca} we outline the main features of the OFSS theory that
describes the behavior of finite systems driven across a classical
first-order transition, and we also discuss the expected behavior in
the TL.  Sec.~\ref{numresofss} reports numerical results for
finite-size systems in two and three dimensions, which support the
OFSS theory presented in Sec.~\ref{outsca}.  In Secs.~\ref{2dTL} and
\ref{3dTL} we study the dynamics in the TL, in two and three
dimensions, respectively.  Finally, in Sec.~\ref{conclu} we summarize
and draw our conclusions. The Appendix presents a coarse-grained
effective model which reproduces the asymptotic OFSS behavior.

\section{Models and dynamic protocol}
\label{moddyn}

\subsection{The Ising model}
\label{isimod}

We consider the 2D, 3D, and 4D nearest-neighbor Ising model in the
presence of a magnetic field $h$, focusing on the dynamic behavior
across the low-temperature first-order transition line.  The
Hamiltonian is
\begin{equation}
 H = - J \sum_{{\bm x},\mu} s_{\bm x} \, s_{{\bm x}+\hat{\mu}} - h
 \sum_{\bm x} s_{\bm x},
\label{eq:isimod}
\end{equation}
where $s_{\bm x}=\pm 1$ are the spin variables associated with the
sites of a cubic-like lattice. The partition function reads
\begin{equation}
  Z = \sum_{\{s_{\bm x}\}} e^{-\beta H},\qquad \beta=1/T.
  \label{partfunc}
\end{equation}
We consider finite systems of size $L$ in each direction with periodic
boundary condition. Moreover, we set $J=1$ without loss of generality.

The Ising model undergoes a continuous transition for $h=0$ at
$\beta_c={1\over 2} \ln(1+\sqrt{2})\approx 0.440687$ in two dimensions
(see, e.g., Refs.~\cite{ID-book1,McCoy-book}), at
$\beta_c=0.221654626(5)$ in three dimensions~\cite{FXL-18}, and at
$\beta_c=0.149693785(20)$ in four
dimensions~\cite{LM-23,LXZFD-24}. Magnetic first-order transitions
occur for $h=0$ and any $\beta > \beta_c$. Along this line the
magnetization,
\begin{equation}
  m(h,\beta) =   {\langle \, \sum_{\bm x} s_{\bm x} \, \rangle\over V},
  \qquad V = L^d,
  \label{magn}
\end{equation}
is discontinuous,
\begin{eqnarray}
\lim_{h\to \pm 0} m(h,\beta>\beta_c) =  \pm \, m_0(\beta).   \label{m0h}
\end{eqnarray}
In two dimensions the magnetization $m_0(\beta)$ is exactly known
(see, e.g., Ref.~\cite{ID-book1} and references therein):
\begin{eqnarray}
m_0(\beta) = \left[ 1 - (\sinh 2\beta)^{-4}\right]^{1/8}.
  \label{m02d}
\end{eqnarray}
For later use, it is also convenient to introduce a rescaled
magnetization defined as
\begin{equation}
  M(h,\beta) = {m(h,\beta)\over m_0(\beta)},
  \label{mrdef}
\end{equation}
so $\lim_{h\to \pm 0} M(h,\beta>\beta_c) =  \pm 1$.
We also consider the bond-energy density defined as
\begin{eqnarray}
  B_e(h,\beta) = - {\langle \sum_{{\bm x},\mu} s_{\bm x} \,
  s_{{\bm x}+\hat{\mu}}\rangle\over V}.
  \label{bondaverage}
\end{eqnarray}

\subsection{The Kibble-Zurek protocol}
\label{KZprot}

We study the out-of-equilibrium dynamics in which the magnetic field
is slowly varied across the first-order transition at fixed $T<T_c$.
We consider specific examples of a purely relaxational dynamics
(model-A dynamics in the standard terminology \cite{HH-77}) and a
dynamic protocol analogous to that proposed by Kibble and Zurek for
the study of the production of defects when slowly crossing continuous
classical and quantum transitions~\cite{Kibble-80,Zurek-96}.  The KZ
protocol that we consider is the following:

\noindent
$\bullet$ The system starts at time $t_i<0$ from 
configurations that are thermalized at a given $\beta>\beta_c$ and 
magnetic field $h_i$. We choose $h_i < 0$, 
so the magnetization is negative at the beginning of the evolution.

\noindent
$\bullet$ The system evolves according to a purely relaxational dynamics
with a time-dependent magnetic field, which varies as
\begin{equation}
  h(t) = {t\over t_s}, 
  \label{htdef}
\end{equation}
where $t_s$ is a fixed time scale,
up to a final time $t_f$. 
The temperature is constant in the evolution.
We set $t_i= t_s \, h_i$  and choose $t_f > 0$ large enough so that
the average magnetization of the configurations obtained at time 
$t = t_f$ is positive. 

Information on the dynamics is provided by the average magnetization
and bond-energy density as a function of time,
\begin{eqnarray}
  &&m(t,t_s,L) = {\langle \,
  \sum_{\bm x} s_{\bm x} \, \rangle_t\over V},\;\;
 M(t,t_s,L) = {m(t,t_s,L)\over m_0},\quad\;\;
  \label{magn2}\\
&&  B(t,t_s,L) = - {\langle \sum_{{\bm x},\mu} s_{\bm x} \,
  s_{{\bm x}+\hat{\mu}}\rangle_t\over V},
  \label{bondaverage2}
\end{eqnarray}
where the average is performed over a large number of trajectories
starting from thermalized configurations at inverse temperature
$\beta$ and $h = h_i$.  In the previous definition, we do not
explicitly report the dependence on $\beta$, which is always kept
fixed in the dynamics (the dependence on $\beta$ is not relevant as
long as $\beta>\beta_c$).

We consider two different dynamic regimes. First, we consider the OFSS
regime, which describes the interplay between the time-dependent
magnetic field $h(t)$ and the finite size $L$, in the limit
$t_s\to\infty$ and $L\to\infty$, keeping some appropriate combinations
of $t$, $t_s$, and $L$ fixed---it will be discussed in
Sec.~\ref{outsca}.  A different out-of-equilibrium scaling regime
occurs in the TL, in which one takes first the infinite-volume limit
keeping the system and protocol parameters fixed, and then one
considers the large-$t_s$ limit.

In our numerical study we consider two different realizations of a
purely relaxation dynamics: a Metropolis local dynamics [where a spin
  flip is performed with probability $P(s_{{\bm x}}\to -s_{{\bm x}}) =
  {\rm Min}(1,e^{-\Delta H})$ at each site, where $\Delta H$ is the
  change of the Hamiltonian when replacing $s_{{\bm x}}$ with $
  -s_{{\bm x}}$] and a heat-bath dynamics (where at each site a new
spin is chosen using the conditional probability distribution at fixed
neighboring spins).  Spins are updated using a checkerboard
scheme. Since cubic-like lattices are bipartite, sites can be divided
in two sets, the set of even and odd sites, respectively. We first
update all spins at even sites, then all spins at odd sites.  A time
unit corresponds to a complete lattice sweep.  Since the Metropolis
and heat-bath dynamics are both representatives of a purely
relaxational dymamics, they are expected to generate analogous
out-of-equilibrium scaling behaviors (see also below).

\section{Out-of-equilibrium scaling behavior}
\label{outsca}

\subsection{Finite-size behavior} \label{OFSS-theory}

In this section we outline the OFSS theory that was originally
proposed in Ref.~\cite{PV-17} for first-order temperature-driven
transitions. An analogous OFSS theory has been proposed for
first-order quantum transitions, see, e.g.,
Refs.~\cite{PV-24,RV-21,PRV-25,PRV-25b}. The OFSS behavior depends on
the nature of the boundary conditions~\cite{PV-24}.  Here, we assume
boundary conditions that preserve translational invariance and do not
favor any of the coexisting phases, as is the case for the periodic
boundary conditions.

To specify the OFSS regime for the KZ dynamics, we should identify
  the appropriate scaling variables.  The first scaling variable is
  $\Phi=h(t) L^d$, which, for time-independent magnetic fields,
  parametrizes the equilibrium finite-size scaling behavior at
  first-order transitions driven by magnetic
  fields~\cite{VRSB-93,PV-24}.  Another scaling variable $\Theta$ can
  be obtained by rescaling the time $t$ with the time scale $\tau(L)$
  of the slowest modes of the dynamics~\cite{PV-17b}.  Therefore, the
  time-dependent large-scale observables are expected to obey OFSS
  laws in terms of the scaling variables~\cite{PV-17,PV-24}
\begin{equation}
  \Phi = h(t) L^d, \qquad \Theta = {t/\tau(L)}.
  \label{rudef}
\end{equation}
OFSS is obtained in the limit $L\to\infty$ and $t,t_s\to\infty$
keeping the scaling variables $\Phi$ and $\Theta$ fixed.  The
identification of $\tau(L)$ requires an understanding of the relevant
mechanism that drives the system from one phase to the other.  We make
the assumption---the numerical analysis presented below will confirm
it---that the relevant mechanism in the OFSS regime is the creation of
strip-like domains parallel to the lattice axes, as it has already
been checked at fixed $h=0$~\cite{BHN-93}.  Note that the creation of
spherical droplets is not relevant in the OFSS limit.  Indeed,
spherical droplets are unstable if they are smaller than a critical
radius $R_c$ of the order of \cite{RTMS-94} $a/h$.  In the OFSS limit
we have $h = \Phi/ L^d$, so $R_c\sim L^d/\Phi$.  Therefore, $R_c \gg
L$ in the large-$L$ limit for $d>1$, implying the irrelevance of the
droplets.

Under the previous assumption, the tunneling time $\tau(L)$ can be
parametrized as
\begin{equation}
  \tau(L) \approx c \, L^\alpha e^{\sigma L^{d-1}}, \qquad 
  \sigma=2\beta\kappa,\
  \label{taul}
\end{equation}
for sufficiently large $L$, where $\kappa$ is the interface free energy.
In two dimensions, $\kappa$ is known exactly ~\cite{ZA-82}:
\begin{equation}
   \kappa = 2 + \beta^{-1}\ln\,\tanh\,\beta.
  \label{sigma}
\end{equation}
The exponent $\alpha$ in Eq.~(\ref{taul}) is instead not known.
Numerical analyses of an equilibrium relaxational dynamics at the
first-order transitions occurring in 2D Ising and Potts systems are all
consistent with 
$\alpha\approx 2$~\cite{PV-17b,PV-17,PV-24}.

The OFSS limit is obtained by taking $t,t_s\to\infty$, and
$L\to\infty$ keeping the scaling variables $\Phi$ and $\Theta$ defined in 
Eq.~(\ref{rudef}) fixed. Equivalently, we can
consider the time-independent scaling variable
\begin{equation}
  \Upsilon = {\Theta\over \Phi} = {t_s\over T(L)},
  \qquad T(L) = L^d\,\tau(L), \quad
\label{wdef}
\end{equation}
where $T(L)$ can be interpreted as the time scale associated with the
passage across the first-order transition point $h=0$.

In the OFSS limit at fixed $\Phi$, $\Theta$, and $\Upsilon$, the
rescaled magnetization is expected to asymptotically behave
as~\cite{PV-24}
\begin{equation}
  M(t,t_s,L) \approx {\cal M}(\Phi,\Upsilon) =
  \widehat{\cal M}(\Phi,\Theta),
  \label{fssmr}
\end{equation}
where ${\cal M}$ and $\widehat{\cal M}$ are scaling functions.  They
are expected to be independent of the temperature (as long as $T<T_c$)
and of other details of the model, provided one fixes the nonuniversal
normalizations of the arguments. In the OFSS limit the bond-energy
density $B$ defined in Eq.~(\ref{bondaverage2}) is expected to scale
as
\begin{eqnarray}
  \Delta B(t,t_s,L) \equiv B(t,t_s,L) - B_e[h(t)] 
 \approx L^{-d} {\cal B}(\Phi,\Upsilon), \quad\;\;
\label{DeltaE-scaling}
\end{eqnarray}
where $B_e(h)$ is the equilibrium value for a magnetic field $h$.

In the adiabatic limit, corresponding to $\Upsilon\to \infty$, the
OFSS function ${\cal M}(\Phi,\Upsilon)$ must approach the static
finite-size scaling function of the rescaled magnetization, which only
depends on $\Phi=h L^d$.  Of course, in the adiabatic limit the OFSS
function associated with $\Delta B(t,t_s,L)$ vanishes.

\subsection{Infinite-volume behavior}

The previous results apply in the OFSS regime, when the time scale
$t_s$ is of the order of the typical transition time $T(L)$ defined in
Eq.~(\ref{wdef}). A different behavior is expected in the TL, in which
we consider the infinite-volume limit at fixed $t_s$, as in this case
no tunneling occurs for $h\approx 0$.  Other mechanisms should play a role
in the TL, such as the nucleation of
droplets~\cite{Binder-87,RTMS-94}, which are not relevant in the OFSS
regime.

\begin{figure}[tbp]
\includegraphics[width=0.9\columnwidth, clip]{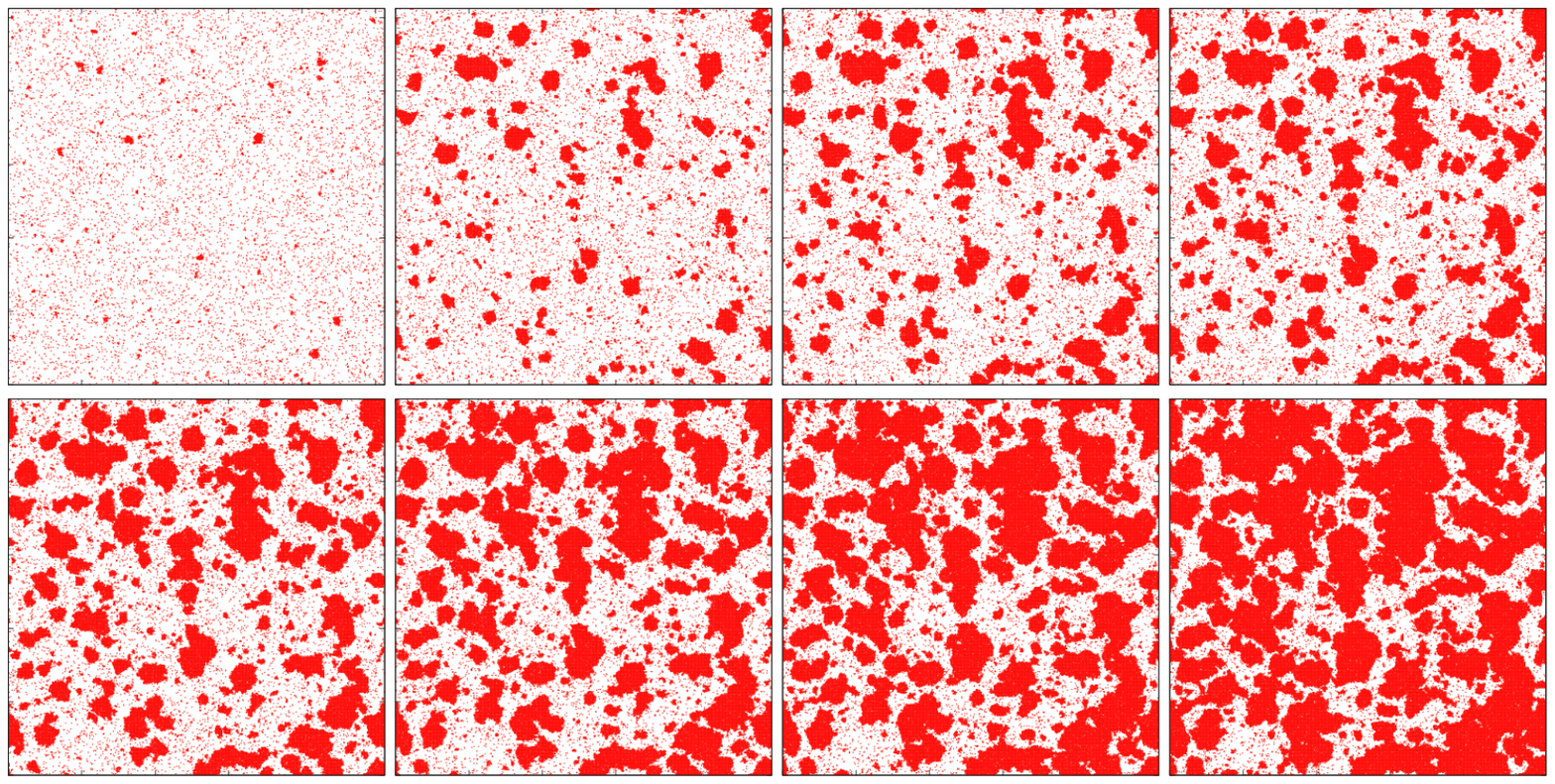}
    \caption{Snapshots of a 2D Ising system of size $L=1024$, for
      different values of $t$. Results for a heat-bath KZ evolution
      with $t_s = 10000$.  A red dot corresponds to a positive spin.
      Time increases moving from left to right along a row.  The
      leftmost top panel shows the system in the metastable state ($m
      \approx -m_0$). The second panel on the left (top row) shows the
      system just before it changes phase (here $m\approx -0.70$).  }
\label{snapshot}
\end{figure}

To identify the appropriate scaling variable for the infinite-size
dynamics, one may assume that the system behaves as a gas of droplets
of size $R\ll L$ for large $L$.  Evidence for this behavior in the 2D
Potts model is provided in Ref.~\cite{PV-17} (see also its
supplementary material). Also numerical data in the 2D Ising model
appear to be consistent with this picture, see Fig.~\ref{snapshot}.
Indeed, the transition to the other phase starts with the nucleation
of droplets of positive spins.  Therefore, the relevant length scale
in the TL should be the time-dependent typical droplet size $R(t)$,
implying that the correct scaling variable is $\Phi = h(t) R(t)^d$. To
completely specify the time dependence of $\Phi$ we should estimate
the typical time needed to create a droplet of size $R$.  If we assume
that the droplet surface is smooth, with an area that scales as
$R^{d-1}$, the time $t$ needed to create a droplet of size $R$ should
increase exponentially with the area $R^{d-1}$ of the droplet, so $\ln
t \sim R^{d-1}$. Thus, we expect $R(t) \sim (\ln t)^{1/(d-1)}$. 
  Under these hypotheses, we end up with the infinite-volume scaling
  variable
\begin{equation}
\widehat\Phi = h(t) (\ln t)^{d/(d-1)} = { t \,(\ln t)^{d/(d-1)} \over t_s}.
\label{rinfdef}
\end{equation}
As we shall see, the numerical analysis of the out-of-equilibrium KZ
dynamics in the TL shows that $\widehat\Phi$ provides the relevant
scaling variable only in two dimensions. As we shall discuss, the
numerical results for the 3D Ising model apparently admit a scaling
description in terms of a different scaling variable. 
Numerical data suggest $t (\ln t)^\kappa$, with $\kappa = 1$. instead  
of the value $\kappa = 3/2$ predicted by Eq.~(\ref{rinfdef}).

\section{Numerical results for the KZ dynamics in the OFSS regime}
\label{numresofss}

\begin{figure}[tbp]
\includegraphics[width=0.9\columnwidth, clip]{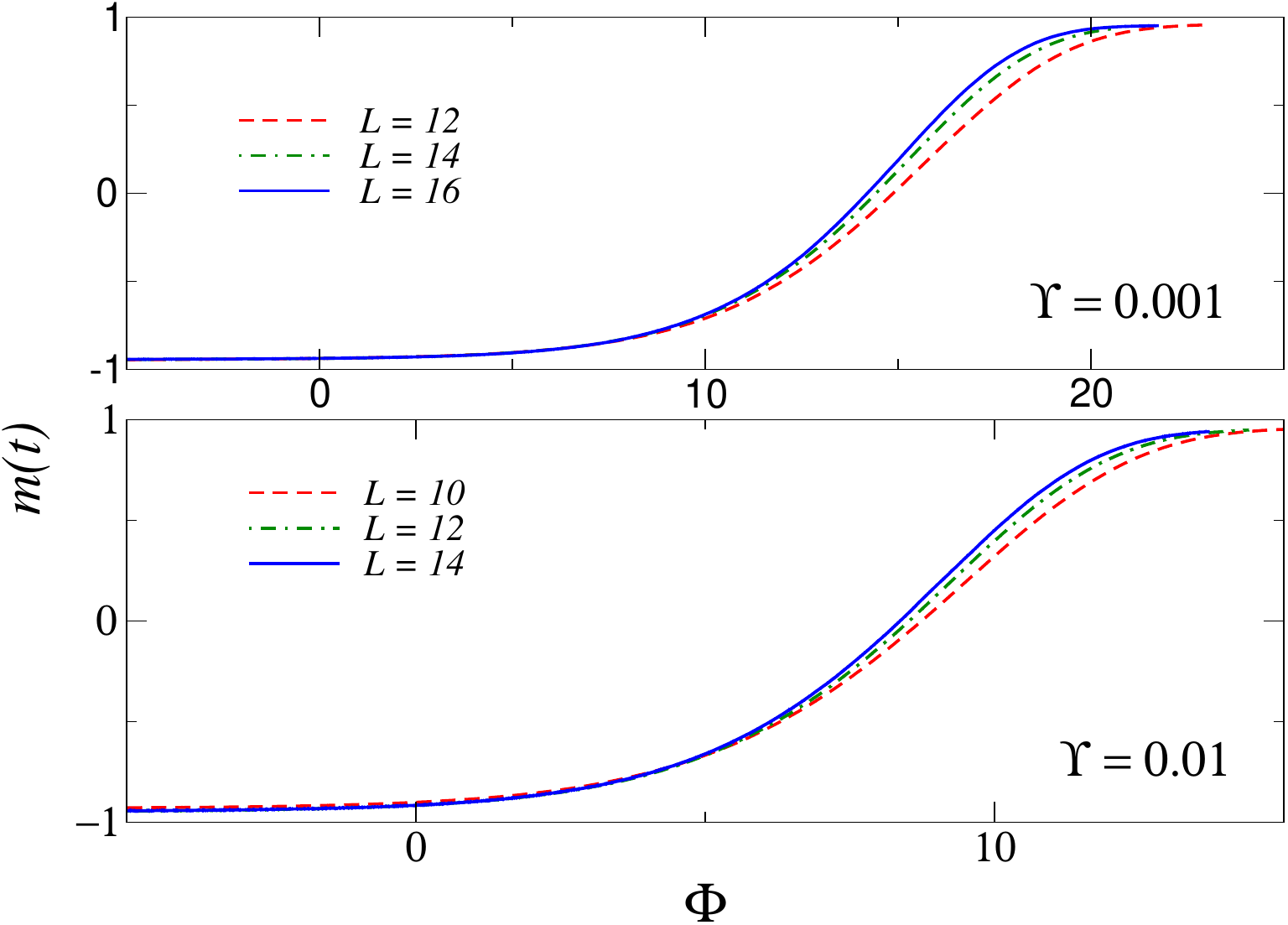}
    \caption{Average time evolution of the magnetization as a function
      of $\Phi =h(t) L^2$. 2D Ising results at $\beta=1.2\beta_c$ for
      $\Upsilon=0.01$ (bottom) and 0.001 (top).  Errors on $m(t)$ are
      at most 0.0002.  }
\label{FSSmagn}
\end{figure}

We now present a numerical analysis of the OFSS behavior in two and
three dimensions, using the heat-bath dynamics.

In two dimensions, we consider a value of $\beta$ deep in the
low-temperature phase, i.e., $\beta = 1.2\beta_c$.  We first verify
Eq.~(\ref{taul}) and estimate the exponent $\alpha$, by performing
some MC simulations at $h=0$.  We compute the average time between two
successive changes of the sign of the magnetization for some values of
$L$. This requires a notable computational effort, because the time
$\tau(L)$ rapidly increases with $L$.  We obtain sufficiently precise
results only for moderately large $L$:
\begin{eqnarray}
  &&\tau(L=10) = 1.17(3) \times 10^5, \label{taulest}\\
  &&\tau(L=12) = 5.4(3) \times  10^5, \nonumber\\
  &&\tau(L=14) = 2.92(8) \times 10^6 .\nonumber
\end{eqnarray}
If we fit the data to Eq.~(\ref{taul}), fixing $\sigma$ to the exact
result reported in Eq.~(\ref{sigma}), we obtain $\alpha =
1.63(11)$. An interpolation of the results for $L=12$ and 14 gives
instead $\alpha = 2.3(4)$. In the following we define the scaling
variable $\Upsilon$, cf. Eq.~(\ref{wdef}), using Eq.~(\ref{taul}) with
$\alpha = 2$ and $c=1$.  Changing $\alpha$ by $\pm 0.4$ does not have
a significant impact on the OFSS analysis.

\begin{figure}[tbp]
  \includegraphics[width=0.9\columnwidth, clip]{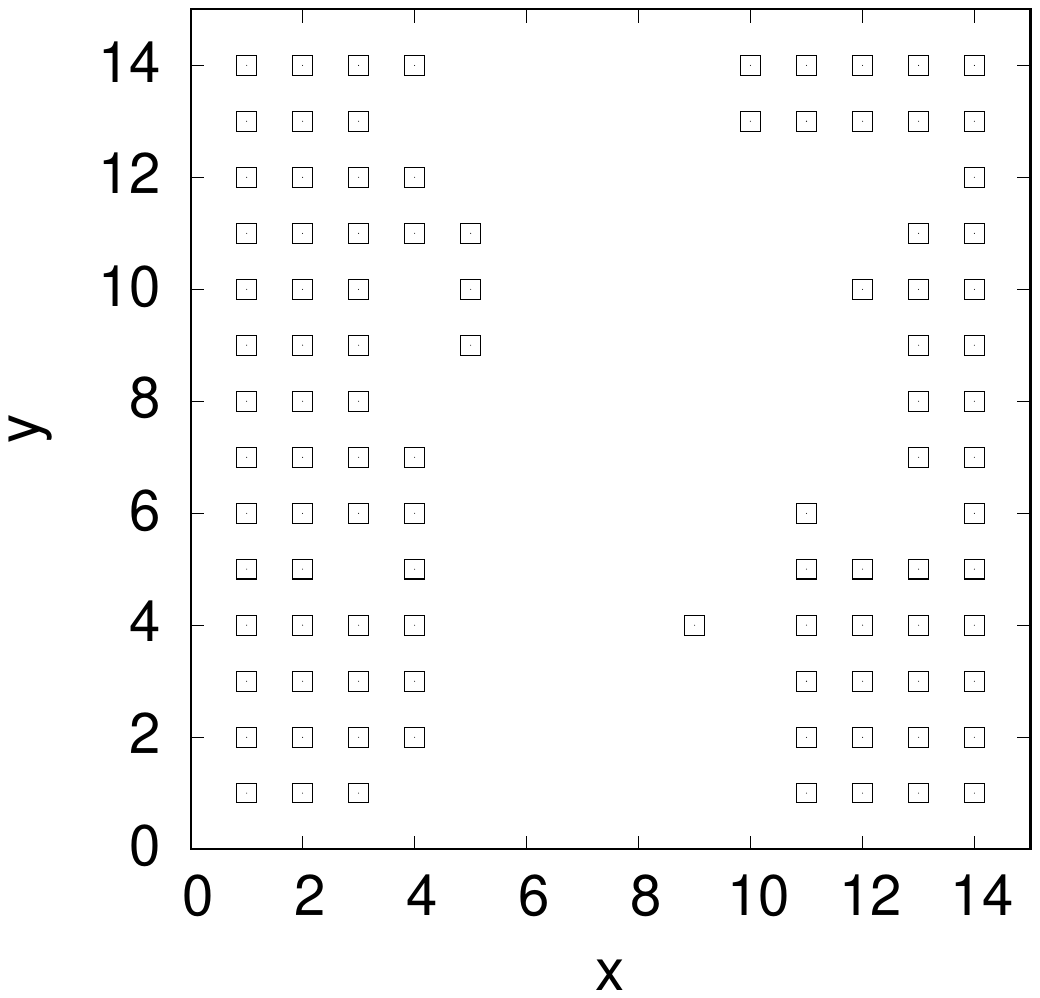}
  \includegraphics[width=0.9\columnwidth, clip]{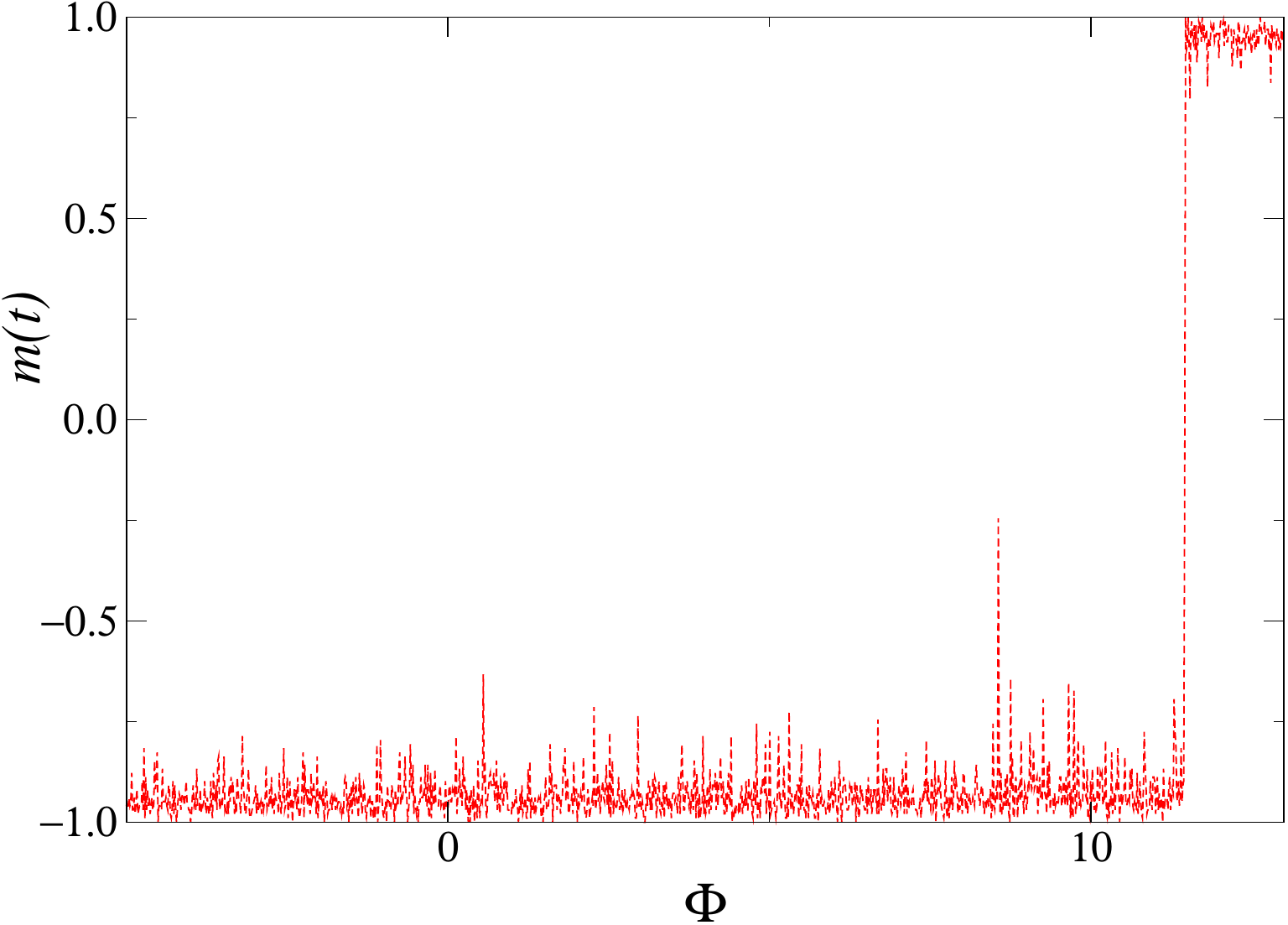}
  \caption{Results for the 2D Ising model at $\beta=1.2\beta_c$.
      Bottom: Plot of the magnetization for one specific heat-bath
      evolution for $L=14$ and $t_s = 4300740$ ($\Upsilon = 0.01$), as
      a function of $\Phi=h(t) L^2$.  Top: Spin configuration when
      the system is changing phase $(M=0)$. Squares label the lattice
      sites where $s_{\bm x} = 1$.  We remind the reader that we are
      using periodic boundary conditions. }
\label{evolution2DL14}
\end{figure}

Figure~\ref{FSSmagn} reports results for the time-dependent
magnetization obtained using the heat-bath relaxational dynamics, for
some values of $\Upsilon$. As $L$ increases, the magnetization data
approach a large-$L$ scaling curve, in agreement with
Eq.~(\ref{fssmr}). Note that scaling corrections increase as
$\Upsilon$ decreases, which is not unexpected, as smaller values of
$\Upsilon$ correspond to smaller values of the time scale $t_s$.

It is important to realize that the reported plots give the behavior
of the average magnetization, the average being taken over a large
number of independent trajectories (it varies between $2\times 10^5$
and $4\times 10^5$).  The behavior of the magnetization for a single
evolution is different. As an example, in the lower panel of
Fig.~\ref{evolution2DL14} we show the magnetization as a function of
the scaling variable $\Phi=h(t) L^2$. It first shows fluctuations
around to $M = -1$ (where $M$ is the rescaled magnetization
$M=m/m_0$), which increase in size as $h(t)$ increases, then it makes
a sudden jump to the opposite-magnetization phase.  The jump is almost
instantaneous (it takes less than 10 lattice sweeps for $L=14$,
therefore a much shorter time than $\tau(L=14)\approx 3\times 10^6$).  In
the upper panel of Fig.~\ref{evolution2DL14} we report the configuration
at the time where $m(t)$ is exactly zero. In agreement with the
discussion reported in Sec.~\ref{outsca}, the configuration is
characterized by two strips, one of negative spins (in the center of
the figure) and one of positive spins across the two vertical
boundaries (we recall that we use periodic boundary conditions).

The behavior observed in Fig.~\ref{evolution2DL14} allows us to obtain
a simple interpretation of the rescaled average magnetization
$M(t)$. If we consider a large number $N_{ev}$ of different
evolutions, at time $t$ the number $N_+$ of systems with positive
magnetization $M\approx 1$ is simply
\begin{equation}
  N_+ = N_{ev} {1 + M(t)\over 2}.
  \label{nplus}
  \end{equation}
  Thus, its time derivative provides the probability (density) that
  the system magnetization switches sign at time $t$. The very simple
  behavior of the magnetization for a single system allows us to
  define a simple effective model that captures the main features of
  the dynamics. Details are reported in the Appendix.

\begin{figure}[tbp]
  \includegraphics[width=0.9\columnwidth, clip]{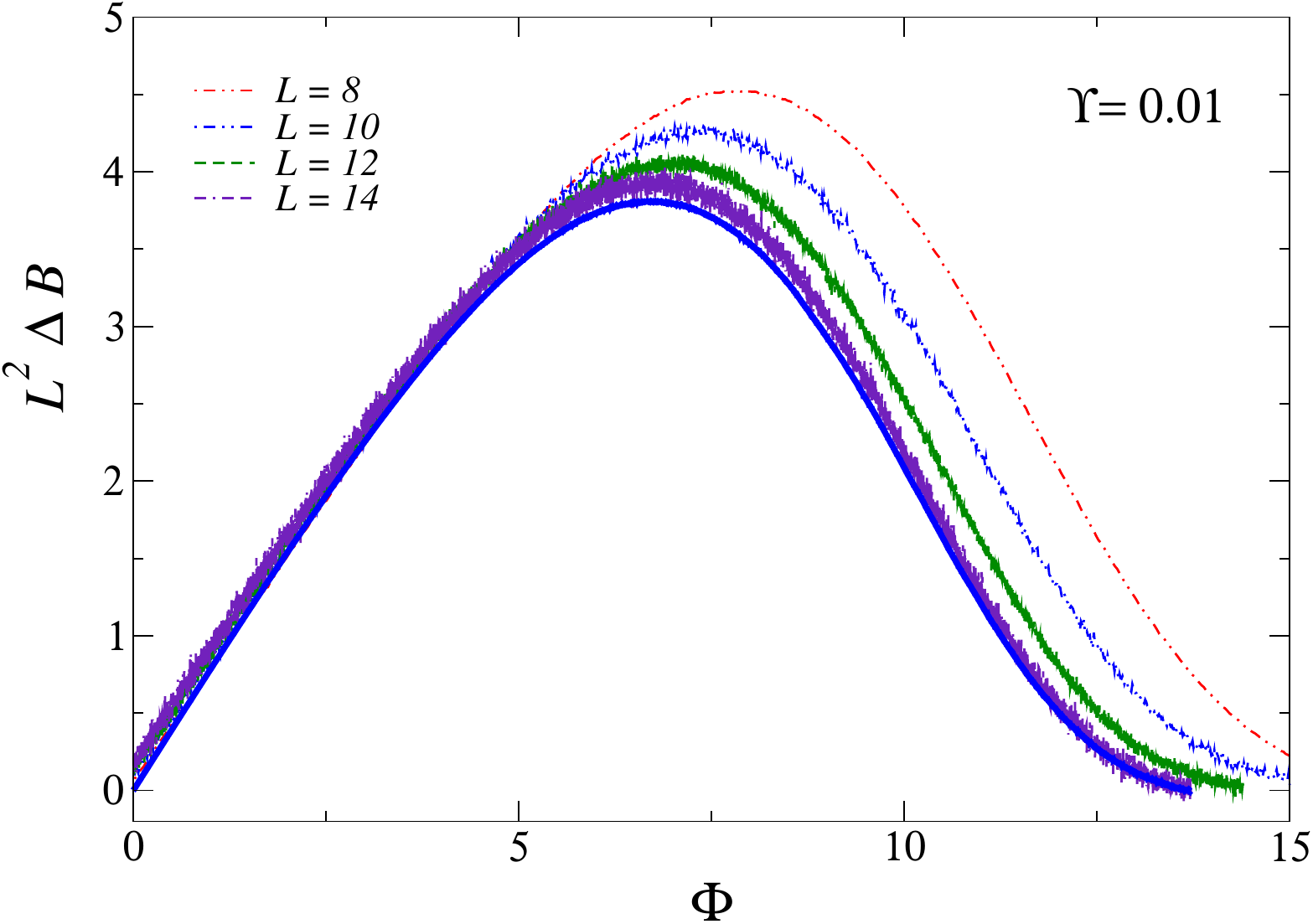}
    \caption{Scaling behavior of the average bond-energy density at
      $\beta=1.2\beta_c$, for $\Upsilon = 0.01$, see
      Eq.~(\ref{DeltaE-scaling}). Results for the 2D Ising model. We
      also report the prediction (blue solid curve) $c \,\Phi [1 -
        M(h)]$, obtained using the $L=14$ data. We set $c = -0.40$
      (obtained by fitting the data). }
\label{FSSenergy}
\end{figure}

We have also analyzed the behavior of the bond-energy density. We have
first determined the equilibrium energy density $B_e(h)$ as a function
of $h$ (by standard MC simulations), and then the energy difference
defined in Eq.~(\ref{DeltaE-scaling}). Data, see Fig.~\ref{FSSenergy},
are consistent with the expected scaling behavior.  The scaling curve
can be predicted by a simple argument.  If $B_{\rm ms}[h(t)]$ is the
bond-energy density of the metastable negative-magnetization state and $[1
  - M(h)]/2$ is the fraction of systems with $M\approx -1$, we have
\begin{equation}
\Delta B(t) = {1 - M[h(t)]\over 2} \Big[B_{\rm ms}[h(t)] -
  B_e[h(t)]\Big].
\end{equation}
The energies have a regular expansion in powers of $h$ and are equal
for $h = 0$, so $B_{\rm ms}(h) - B_e(h) \propto h = (h L^2)/L^2$: the
difference behaves as $1/L^2$ in the scaling limit. Thus, $L^2 \Delta
B(t) = c \, \Phi [1 - M(h)]$, where $c$ is a constant.  This prediction
is in full agreement with the data, see Fig.~\ref{FSSenergy}.

The analysis of the OFSS regime in three dimensions is much more
difficult than in two dimensions. First, $\tau(L)\sim e^{\sigma L^2}$
increases very rapidly with $L$, so, at fixed $\Upsilon$, one ends up
with very large values of $t_s$ even for relatively small values of
$L$.  Second, in the absence of exact results for the interface free
energy, we have been forced to estimate $\tau(L)$ numerically, which
requires very long simulations.  Therefore, we have been only able to
study the dynamics for small values of $L$ ($L\le 11$) and not too far
from the critical point $\beta_c\approx 0.222$ ($\sigma$ increases
rapidly as the temperature decreases, as we show below).  Results for
$\tau(L)$ for two close values of $\beta$ ($\beta = 0.240$ and 0.242)
are reported in Table~\ref{table-tau}. As expected, $\tau(L)$
increases exponentially with $L^2$, and we estimate $\sigma =
0.11(1)$ and 0.12(1) for $\beta = 0.240$ and 0.242. Note that
$\tau(L)$ for fixed values of $L$ increases significantly with
increasing $\beta$, forbidding us from considering smallest values of
the temperature.

\begin{table}
\caption{Estimates of the average time $\tau(L)$ (measured in lattice
  sweeps) between two successive changes of the sign of the
  magnetization. Results for the 3D Ising model (heat-bath dynamics)
  for $h=0$ and two values of $\beta$. Results are consistent with
  Eq.~(\ref{taul}), taking $\sigma = 0.11(1)$ and 0.12(1) for $\beta =
  0.240$ and 0.242, respectively.}
\label{table-tau}
\begin{tabular}{lcc}
\hline\hline 
$\qquad$ &  $\beta = 0.240$ & $\beta = 0.242$ \\
\hline
$L=8$ & 6577(11) & 13765(34) \\
$L=9$ & 39600(100) & 106690(460) \\
$L=10$ & 306960(700) & $[1.115(5)]\times 10^6$ \\
$L=11$ & $[3.004(17)]\times 10^6$  & $[15.72(21)]\times 10^6$ \\
\hline\hline
\end{tabular}\end{table}\

\begin{figure}[tbp]
\includegraphics[width=0.9\columnwidth, clip]{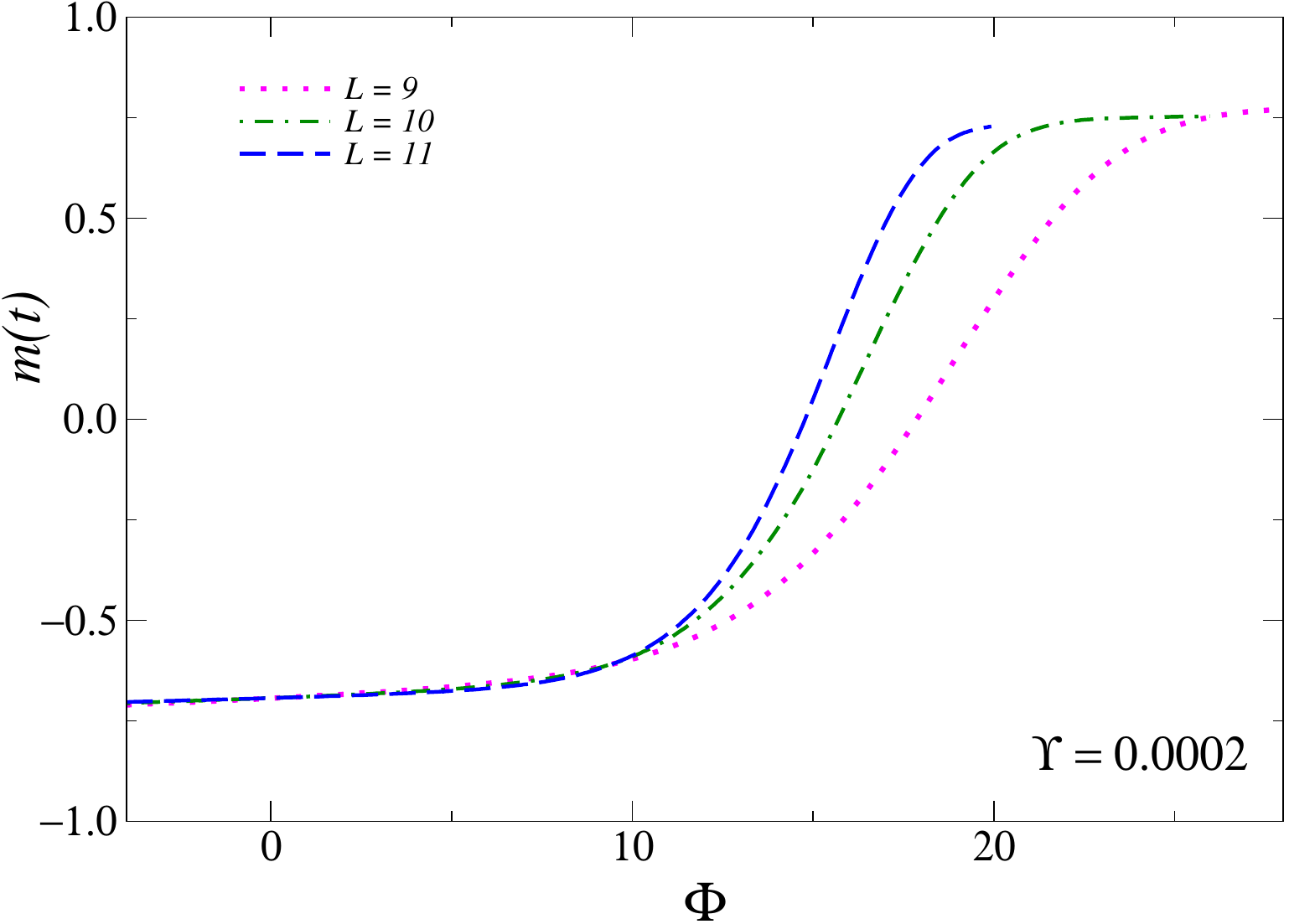}
    \caption{Average time evolution of the magnetization as a function
      of $\Phi =h(t) L^3$. Results for the 3D Ising model (heat-bath
      dynamics) at $\beta=0.242$ for $\Upsilon=0.0002$.  }
\label{FSSmagn3d}
\end{figure}

We analyze the heat-bath KZ dynamics for the largest value of $\beta$,
$\beta = 0.242$, considered above, as we expect convergence to worsen
as the critical point is approached.  The results for $\Upsilon =
0.0002$ reported in Fig.~\ref{FSSmagn3d} are substantially consistent
with the OFSS predictions: As $L$ increases the different curves get
closer, consistently with the existence of a limiting curve for $L\to
\infty$. However, larger lattice sizes are necessary to achieve a
compelling evidence.

\section{The 2D thermodynamic limit}
\label{2dTL}

In this section, and in the following one, we analyze the out-of-equilibrium
behavior of the Metropolis dynamics in the TL, defined as the limit
$L\to \infty$ keeping the Hamiltonian and protocol parameters fixed.
Here, we consider the 2D model and we study the KZ dynamics at fixed
$\beta$---we consider $\beta = 1.2\,\beta_c \approx 0.528824$ and
$\beta=1.1\,\beta_c\approx 0.484755$---starting from $h_i=-0.01$ in
all cases (the actual value of $h_i<0$ is not relevant, since the
out-of-equilibrium scaling behavior is independent of it).  For each
$t_s$ we have performed simulations for several values of $L$,
increasing $L$ until the average magnetization curves become
approximately $L$-independent. Therefore, the curve for the largest
value of $L$ provides the infinite-size limiting curve for the given
values of $t_s$.  We consider several values of $t_s$ to determine the
out-of-equilibrium scaling behavior in the large-$t_s$ limit.  The
time-dependent magnetization is averaged over a large number of
independent trajectories, which varies typically between 20000 (for
the largest systems) and a few millions.

\begin{figure}[tbp]
  \includegraphics[width=0.9\columnwidth, clip]{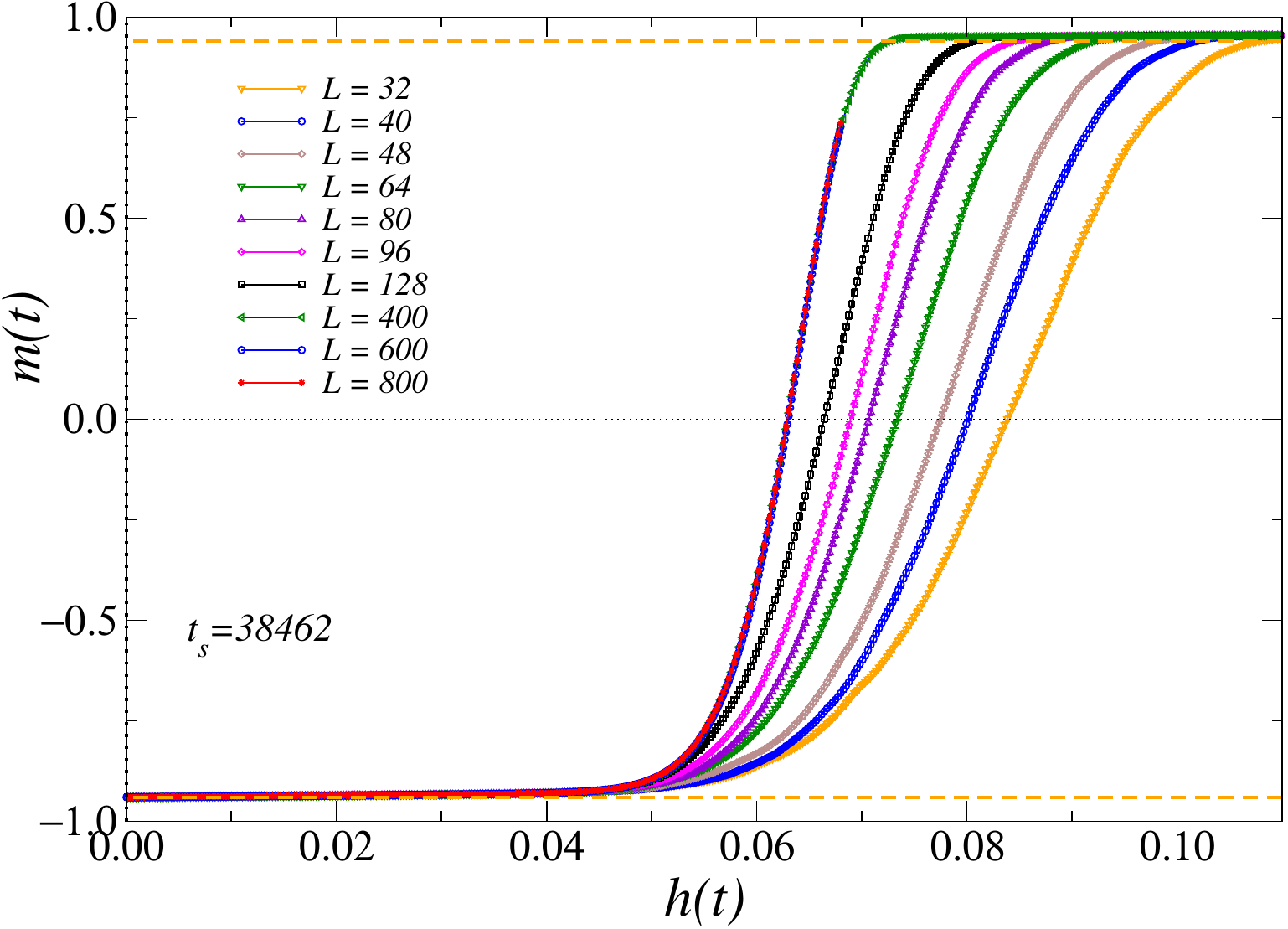}
  \caption{The time evolution of the magnetization $m(t,t_s,L)$ versus
    $h(t)=t/t_s$ for $t_s=38462$ and several sizes $L$. Results for
    the 2D Ising model and the Metropolis dynamics.  Statistical
    errors are hardly visible on the scale of the figure.  Data
    converge to an infinite-size limiting curve as $L$ increases, as
    shown by the agreement within errors of the data for $L\ge 400$.
    The dashed lines correspond to $m=\pm m_0$, where $m_0=0.940259$
    is the equilibrium magnetization at $h=0^\pm$, see
    Eq.~(\ref{m02d}).}
\label{rawlts}
\end{figure}

\begin{figure}[tbp]
  \includegraphics[width=0.9\columnwidth, clip]{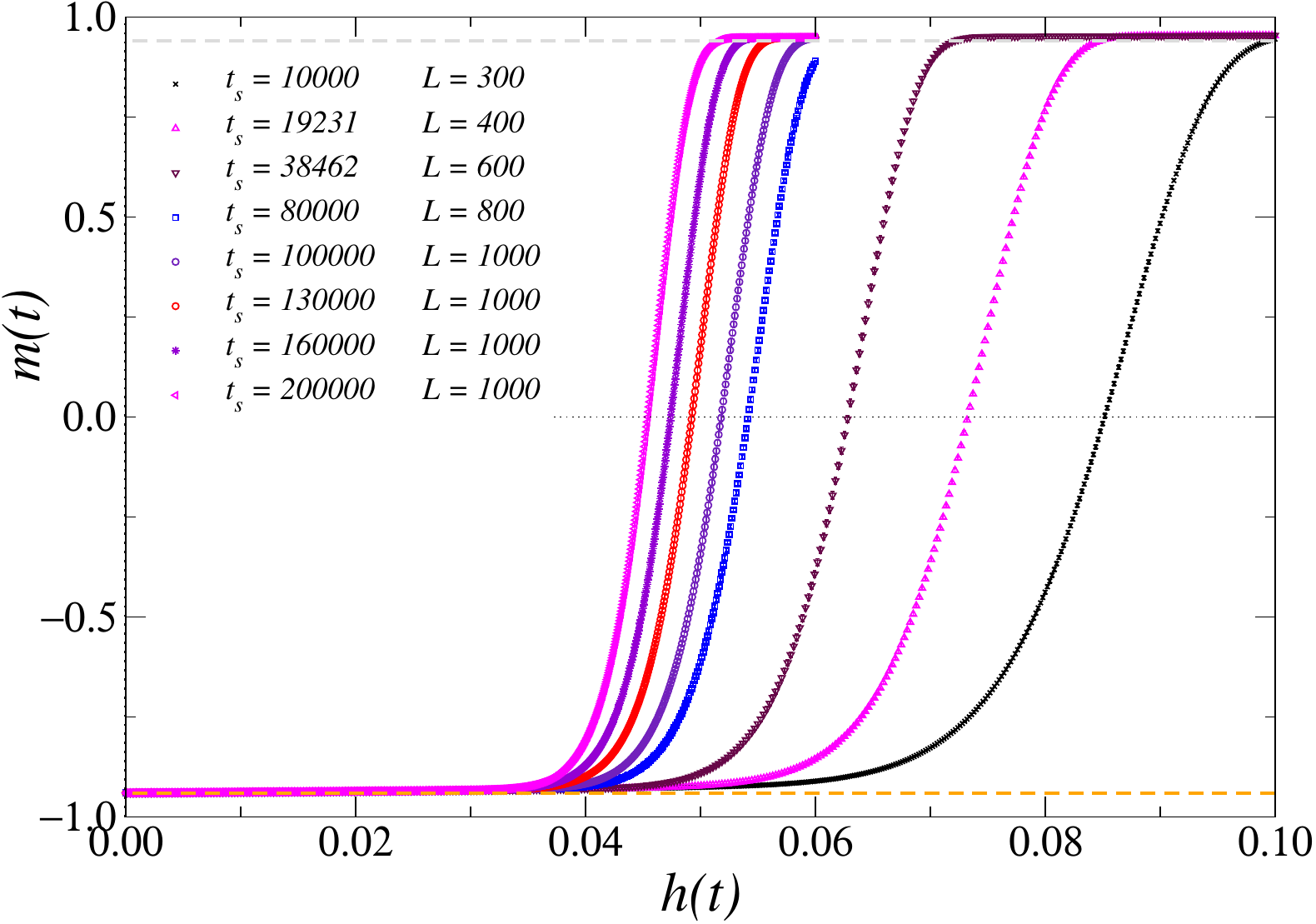}
  \caption{ The time-dependent average magnetization $m(t,t_s,L)$
    versus $h(t)$ for several values of $t_s$ and $L$, at fixed
    $\beta=1.2\beta_c$. Results for the 2D Ising model and the
    Metropolis dynamics.  In all cases, the size $L$ is the largest
    one we have considered for the given value of $t_s$ and is such
    that $m(t,t_s,L)$ can be considered as the average magnetization
    in the TL.  Statistical errors are hardly visible on the scale of
    the figure.  The dashed lines correspond to $m=\pm m_0$, where
    $m_0=0.940259$ is the equilibrium magnetization at $h=0^\pm$.}
\label{rawts}
\end{figure}

\begin{figure}[tbp]
  \includegraphics[width=0.9\columnwidth, clip]{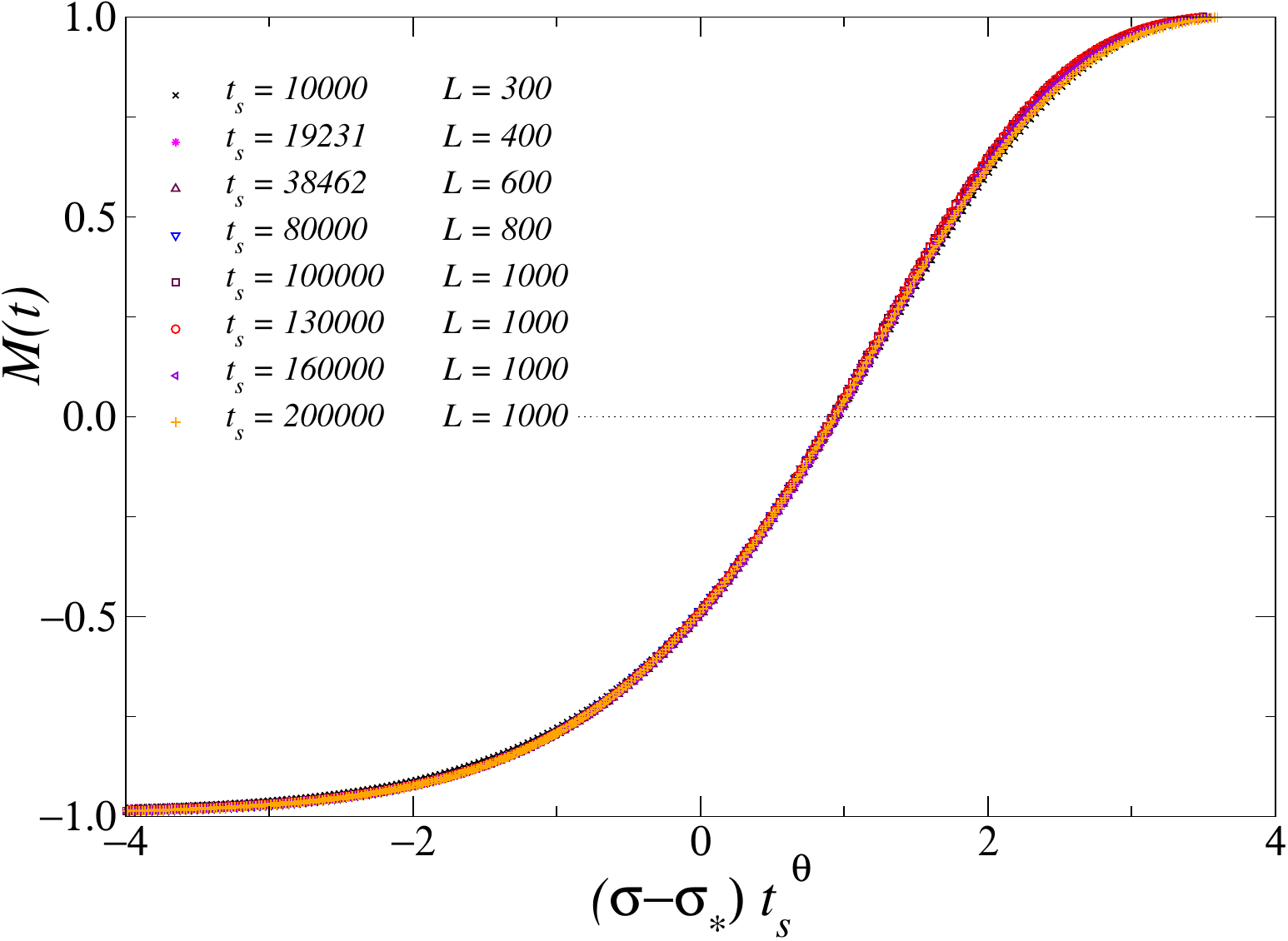}
  \includegraphics[width=0.9\columnwidth, clip]{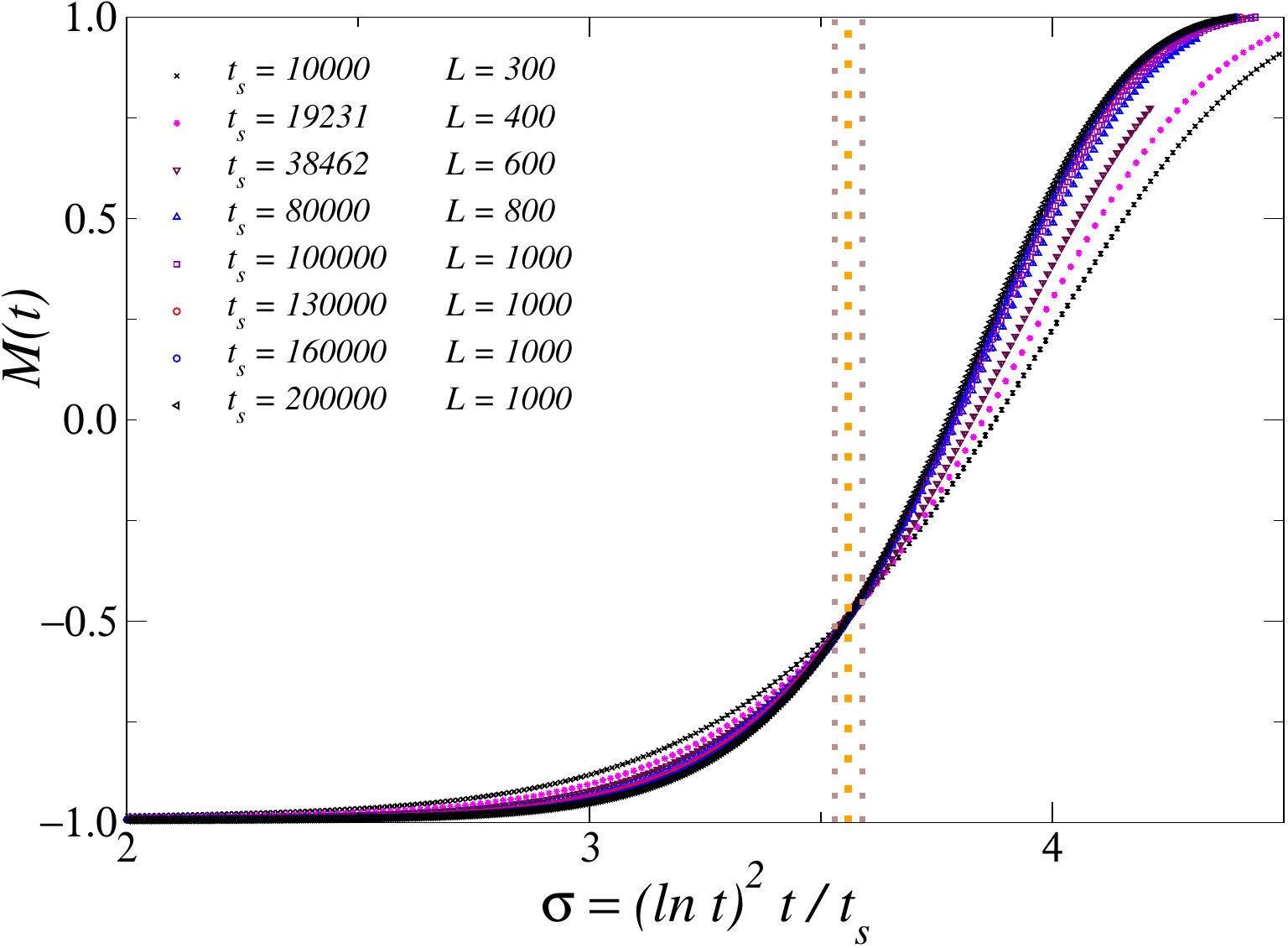}
  \caption{Data for the rescaled magnetization $M = m/m_0$ in the
    TL. Metropolis-dynamics results for $\beta=1.2\beta_c$ in two
    dimensions.  In the lower panel, data are plotted versus $\sigma =
    h(t)(\ln t)^2$; they show a crossing point at
    $\sigma=\sigma_*\approx 3.56$ (the vertical dashed lines
    correspond to $\sigma=3.56$ with an uncertainty of $\pm 0.03$).
    In the upper panel, data are plotted versus
    $\hat\sigma=(\sigma-\sigma_*) t_s^\theta$ with $\theta=0.12$.  }
\label{resc}
\end{figure}

In Fig.~\ref{rawlts} we report results for the Metropolis KZ dynamics
at fixed $\beta=1.2\,\beta_c$. We show the average magnetization for
$t_s=38462$ and several values of $L$. For small values of $L$ we
observe a drift in the magnetization curves as $L$ increases. For
instance, the magnetization changes sign at values of $t$ that
decrease with increasing $L$. The drift of the curves decreases as
largest sizes are considered, and, for $L\ge 400$, the different
curves fall one on top of the other (within errors), providing an
accurate estimate of the time-dependent average magnetization in the
TL.  Analogous results are obtained for all values of $t_s$
considered. In particular, convergence within errors is observed when
$L$ satisfies the approximate inequality $L \gtrsim 2
\sqrt{t_s}$. Therefore, numerical simulations on systems with $L\le
1000$ allow us to accurately determine the infinite-volume
magnetization curves for $t_s\lesssim 2\times 10^5$. In
Fig.~\ref{rawts} we report the estimates of the infinite-size
magnetization versus $h(t)=t/t_s$, for several values of $t_s$.

\begin{figure}[htbp]
\includegraphics[width=0.9\columnwidth, clip]{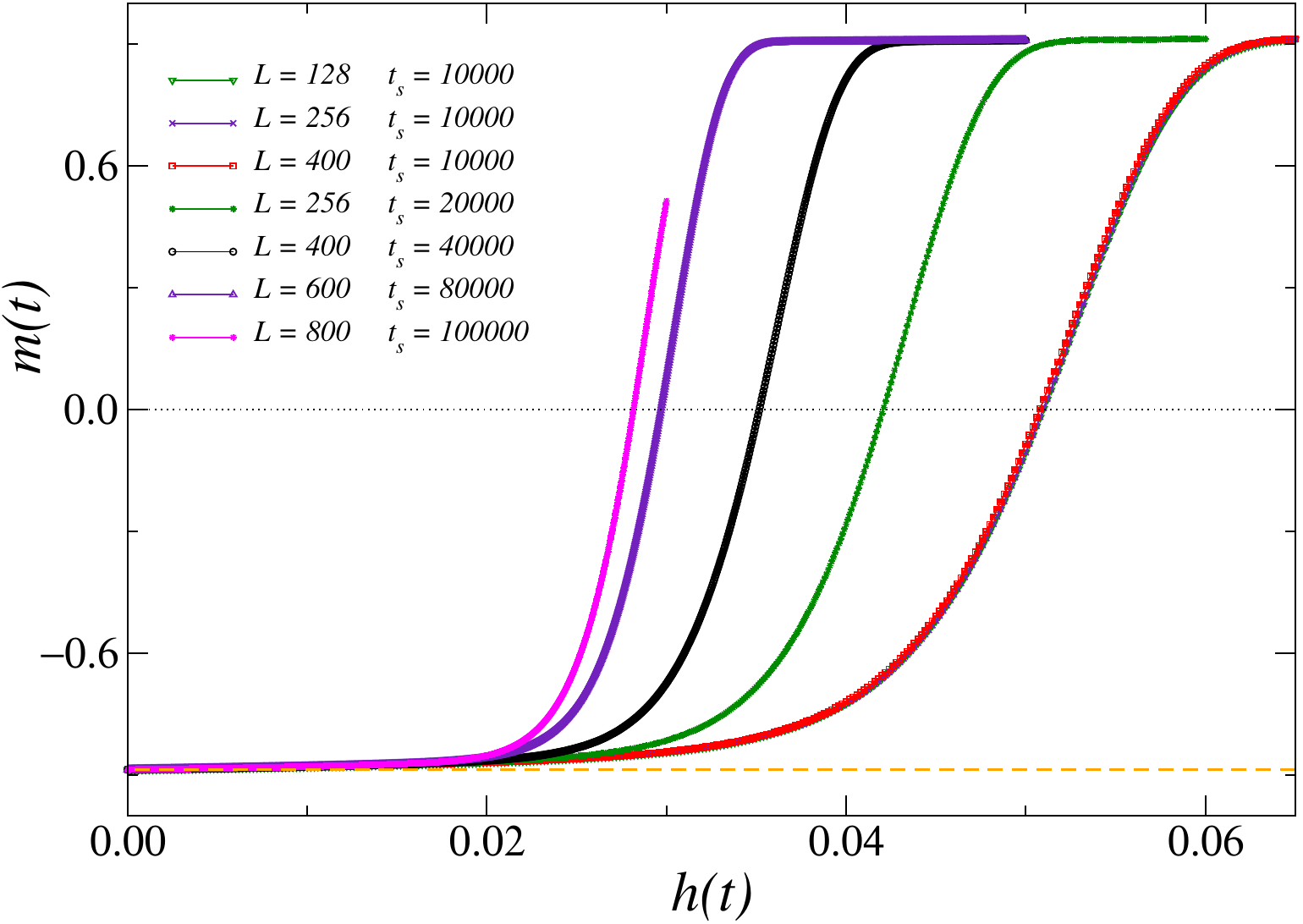}
\caption{Data for the magnetization in the TL for $\beta=1.1\,\beta_c$
  (2D Ising model with Metropolis dynamics).  Data for $m(t,t_s,L)$
  versus $h(t)$ for several values of $t_s$ and values $L$ such that
  convergence to the infinite-size limit has been achieved; the dashed
  line corresponds to the equilibrium magnetization $m=-m_0=-
  0.887193$ for $h=0^-$; analogous data for $\beta = 1.2\beta_c$ are
  reported in Fig.~\ref{rawts}.  }
\label{datasetbeta2a}
\end{figure}

\begin{figure}[htbp]
\includegraphics[width=0.9\columnwidth, clip]{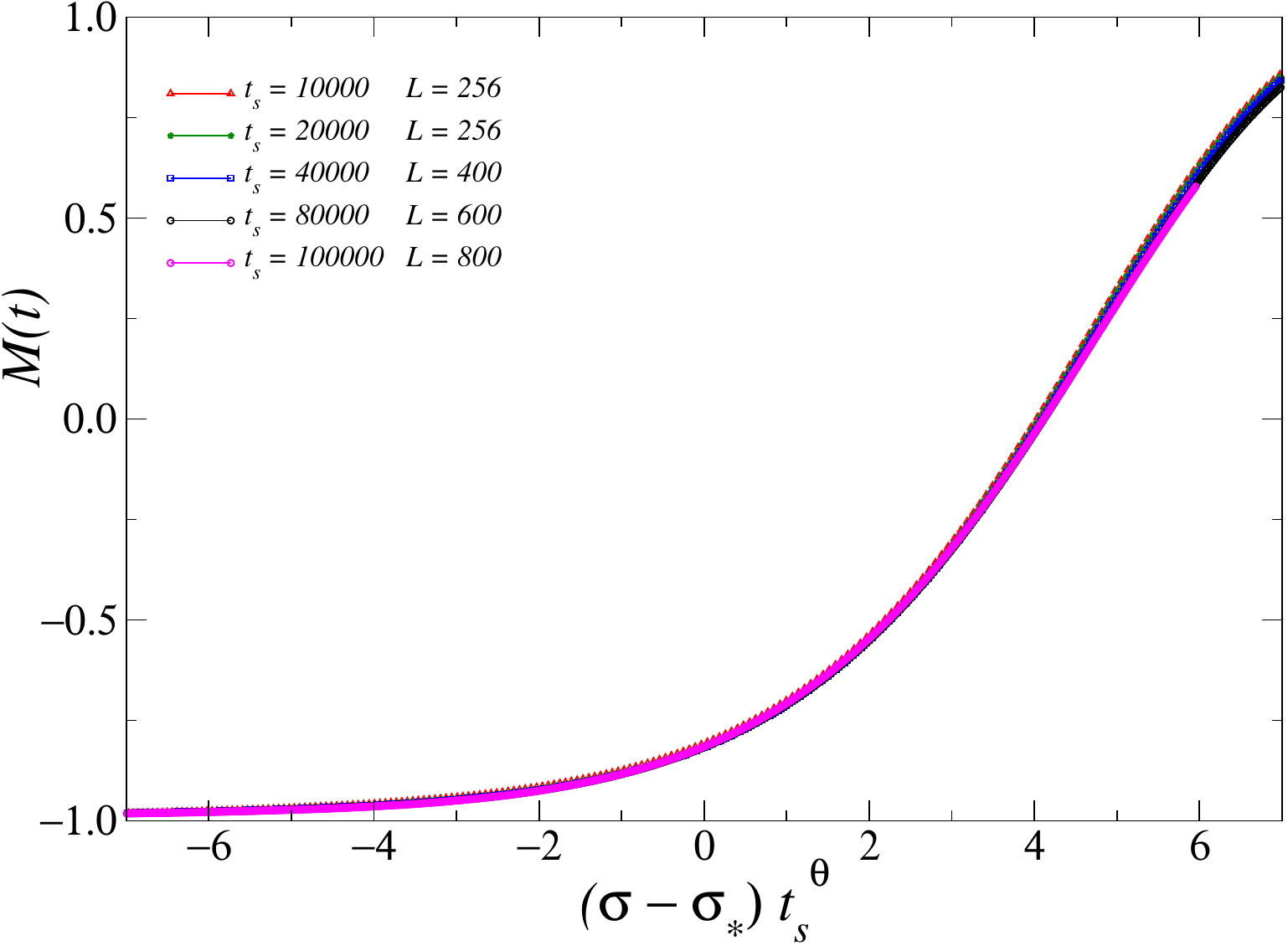}
\includegraphics[width=0.9\columnwidth, clip]{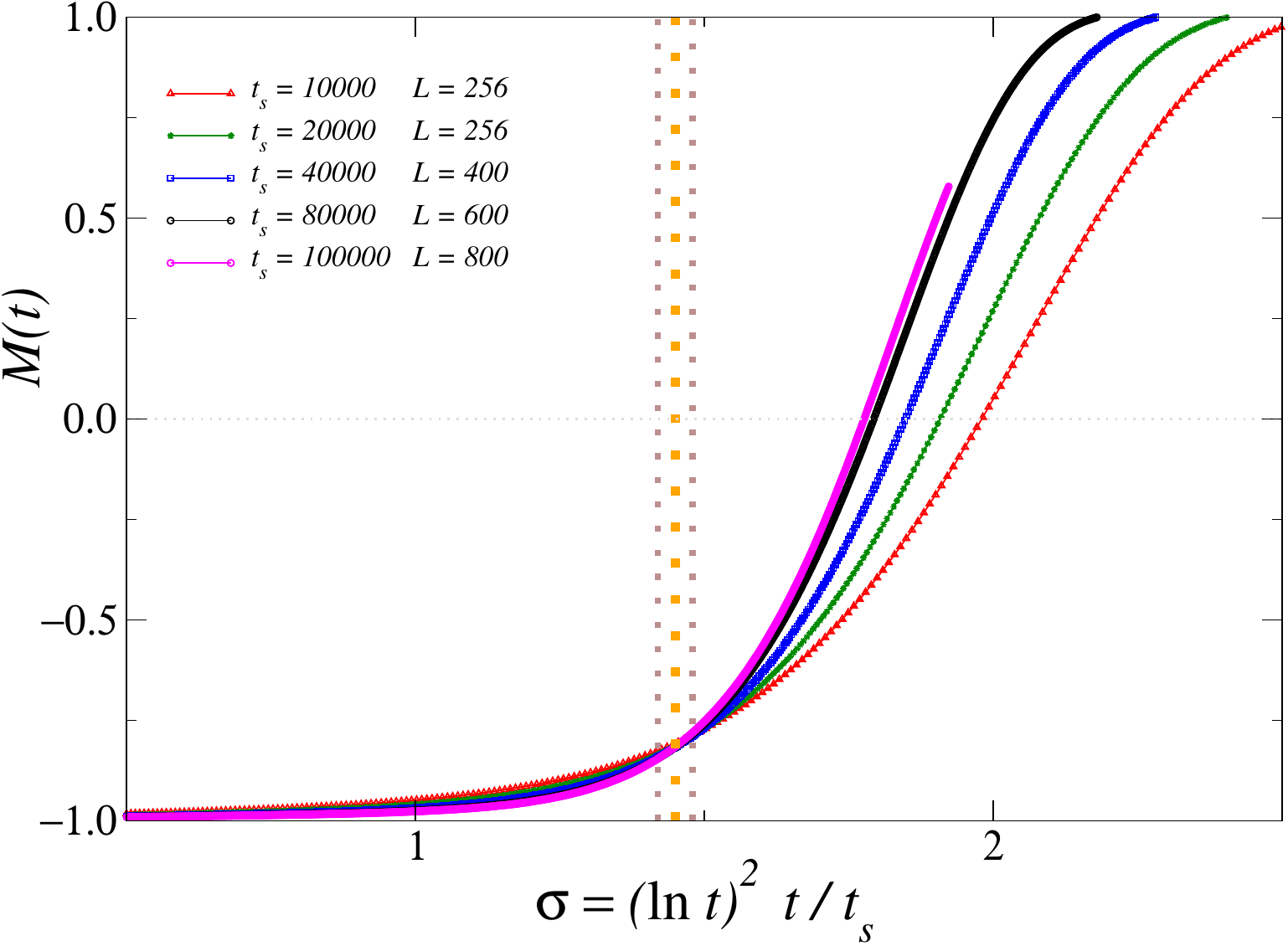}
\caption{Data for the magnetization in the TL for $\beta=1.1\,\beta_c$
  (2D Ising model with Metropolis dynamics).  Top: rescaled
  magnetization $M$ versus $\hat{\sigma}=(\sigma-\sigma_*) t_s^\theta$
  with $\theta=0.22$.  Bottom: rescaled magnetization $M=m/m_0$ versus
  $\sigma$; it shows a crossing point at $\sigma=\sigma_*\approx
  1.45$.  See Fig.~\ref{resc} for analogous results for $\beta = 1.2\,
  \beta_c$.  }
\label{datasetbeta2b}
\end{figure}

In the lower panel of Fig.~\ref{resc} we plot the rescaled magnetization
$M=m/m_0$, versus the scaling variable
\begin{equation}
\sigma = { t \,(\ln t)^2 \over t_s},
\label{sdef}
\end{equation}
which has been defined in Sec.~\ref{outsca}, assuming the system to
change phase by nucleating smooth spherical droplets.  The different
magnetization curves get closer and closer as a function of
$\sigma$. Moreover, they cross at $\sigma=\sigma_*\approx 3.56$, where
the magnetization takes the value $M=M_*\approx -0.5$.  An analogous
behavior was observed at the thermal first-order transitions of the 2D
Potts models~\cite{PV-17}, where it was interpreted as the emergence
of an asymptotic discontinuity of the scaling functions in the
large-$t_s$ limit, whose approach is controlled by a new exponent
$\theta$.  We conjecture an analogous behavior for the time dependence
of the magnetization at the first-order magnetic transitions in 2D
Ising systems, described by the scaling Ansatz
\begin{equation}
M_\infty(t,t_s) \approx \widehat{\cal M}_\infty(\hat{\sigma}), \quad
\hat{\sigma} = (\sigma-\sigma_*) \, t_s^\theta,
\label{estar}
\end{equation}
close to the crossing point, where $\theta>0$ is an appropriate
exponent. As shown in the upper panel of Fig.~\ref{resc}, the Ansatz
(\ref{estar}) provides an excellent parametrization of the
magnetization data close to $\sigma_*$.  Fits of the data to the
ansatz (\ref{estar}) lead to the estimates
\begin{eqnarray}
  \sigma_*=3.56(3),\qquad \theta = 0.12(2),
  \label{s2startheta}
\end{eqnarray}
where the errors take into account the variation of the results when
the data for the smallest values of $t_s$ are systematically excluded
and when the interval of $\sigma$ allowed in the fits is
varied. Statistical errors are negligible.  The scaling behavior
(\ref{estar}) implies that, in the limit $t_s\to\infty$ at fixed
$\sigma$, the magnetization is discontinuous at the crossing point
$\sigma_*$: The rescaled magnetization approaches $M\to -1$ and $M\to
+1$ for $\sigma< \sigma_*$ and $\sigma>\sigma_*$, respectively, for
$t_s\to\infty$.  Corrections decay slowly, as $t_s^{-\theta}$ for
$t_s\to\infty$.

\begin{figure}[tbp]
  \includegraphics[width=0.9\columnwidth, clip]{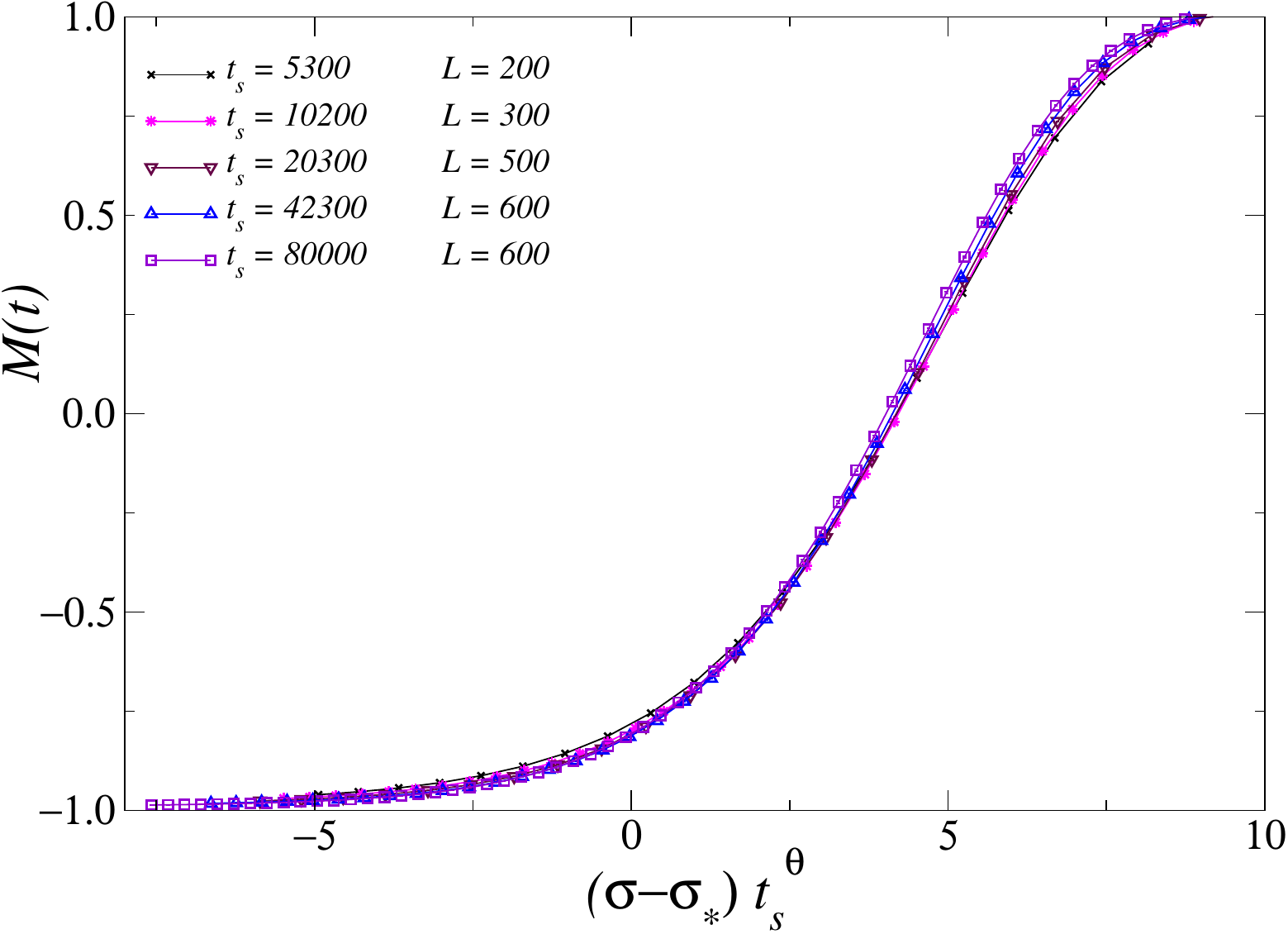}
  \includegraphics[width=0.9\columnwidth, clip]{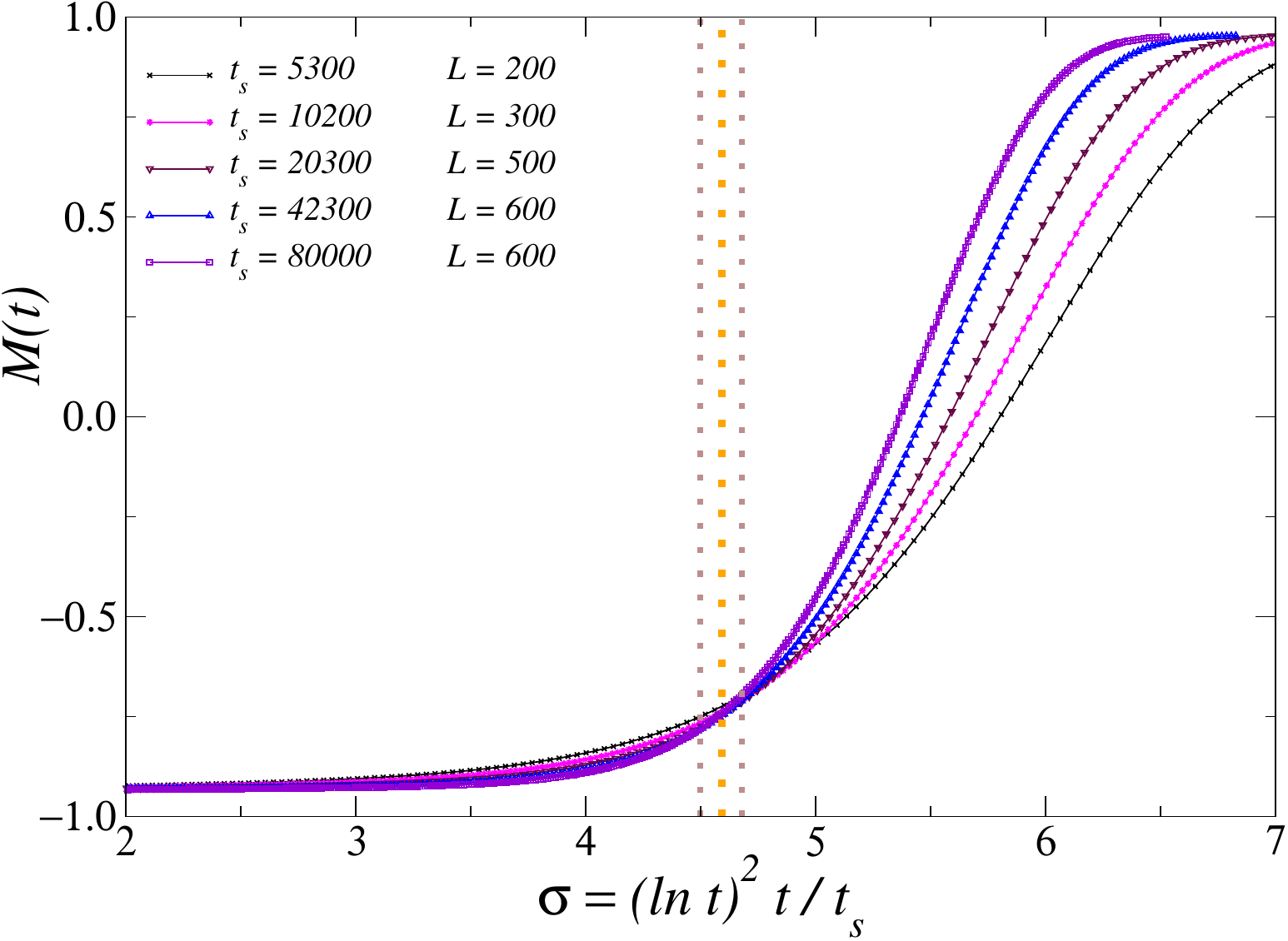}
  \caption{Data for the rescaled magnetization $M = m/m_0$ in the TL
    for a heat-bath dynamics.  Results for $\beta=1.2\beta_c$ in two
    dimensions.  In the lower panel, data are plotted versus $\sigma =
    h(t)(\ln t)^2$; they show a crossing point at
    $\sigma=\sigma_*\approx 4.59$ (the vertical dashed lines
    correspond to $\sigma=4.59$ with an uncertainty of $\pm 0.09$).
    In the upper panel, data are plotted versus
    $\hat\sigma=(\sigma-\sigma_*) t_s^\theta$ with $\theta=0.145$.
    Metropolis results for the same temperature are shown in
    Fig.~\ref{resc}. }
\label{rescHB}
\end{figure}

A similar behavior is obtained for the Metropolis dynamics at
$\beta=1.1\,\beta_c\approx 0.484755$, as shown in
Figs.~\ref{datasetbeta2a} and~\ref{datasetbeta2b}.  Again the
infinite-size data for the magnetization scale when plotted versus
$\hat{\sigma} = (\sigma-\sigma_*) \, t_s^\theta$, with the optimal
values $\sigma_*=1.45(3)$ and $\theta=0.22(2)$.  However, the exponent
$\theta$ and the scaling function differ from those found at
$\beta=1.2\,\beta_c$. In particular, at
the crossing point $\sigma=\sigma_*$ ($\sigma_*\approx 3.56$ and
$\sigma_*\approx 1.45$ for $\beta=1.2\,\beta_c$ and
$\beta=1.1\,\beta_c$, respectively), we have $M=M_*\approx -0.5$ for
$\beta=1.2\,\beta_c$, and $M=M_*\approx -0.8$ for
$\beta=1.1\,\beta_c$.  Therefore, there is no universality with
respect to the temperature.

Previous results have been obtained using the Metropolis dynamics. We
have repeated the same analysis using the heat-bath dynamics at $\beta
= 1.2\beta_c$. Again, the magnetization data scale according with
Eq.~(\ref{estar}), if one uses $\theta = 0.145(20)$ and $\sigma_* =
4.59(9)$, which should be compared with the Metropolis results at the
same temperature, $\theta = 0.12(2)$ and $\sigma_* = 3.56(3)$.  As
shown in Fig.~\ref{rescHB}, scaling is excellent.  As expected,
$\sigma_*$ is larger since the heat-bath dynamics is slower that the
Metropolis one.\footnote{In Ising systems, the heat-bath dynamics is a
Metropolis-Hastings \cite{Hastings-70} dynamics with an acceptance
rate that is smaller than in the standard Metropolis dynamics.} The
value of $\theta$ is instead close to the Metropolis value. This does
not however imply universality, as the scaling functions differ. In
the heat-bath case $M\approx -0.8$ at the crossing point, to be
compared with $M\approx -0.5$ for the Metropolis dynamics.

The value $\sigma_*$, where the magnetization curve is singular,
corresponds to a $t_s$-dependent magnetic field
\begin{equation}
h_* \sim (\ln t_s)^{-2},
\label{hs0}
\end{equation}
which converges to zero when $t_s\to \infty$.  The observed scaling
behavior resembles the one predicted at first-order transitions by
mean-field calculations~\cite{Binder-87}.  There is, however, a
crucial difference: In the mean-field case the singular magnetic
field---the so-called spinodal point--- is different from zero, while
here $h_*$ vanishes for $t_s\to \infty$.  We will refer to the scaling
behavior observed in our case as a spinodal-like behavior.

We finally mention that an analogous scaling behavior was observed at
the thermal first-order transitions in $q$-state 2D Potts
models~\cite{PV-17}.  In particulat, for $q=20$ data are consistent
with the scaling behavior (\ref{estar}) with $\theta \approx 1/3$.

\begin{figure}[tbp]
    \includegraphics[width=0.9\columnwidth, clip]{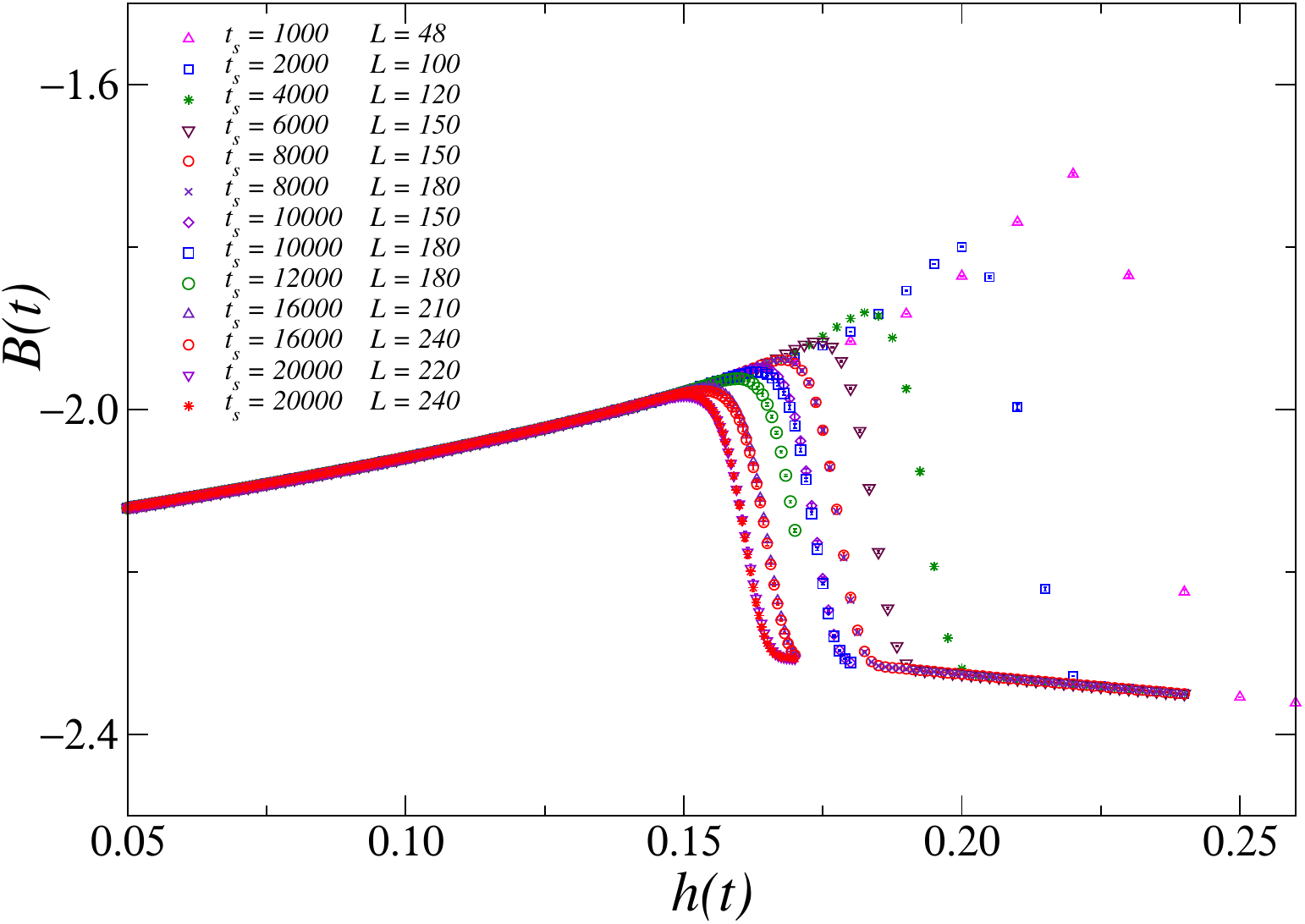}
    \includegraphics[width=0.9\columnwidth, clip]{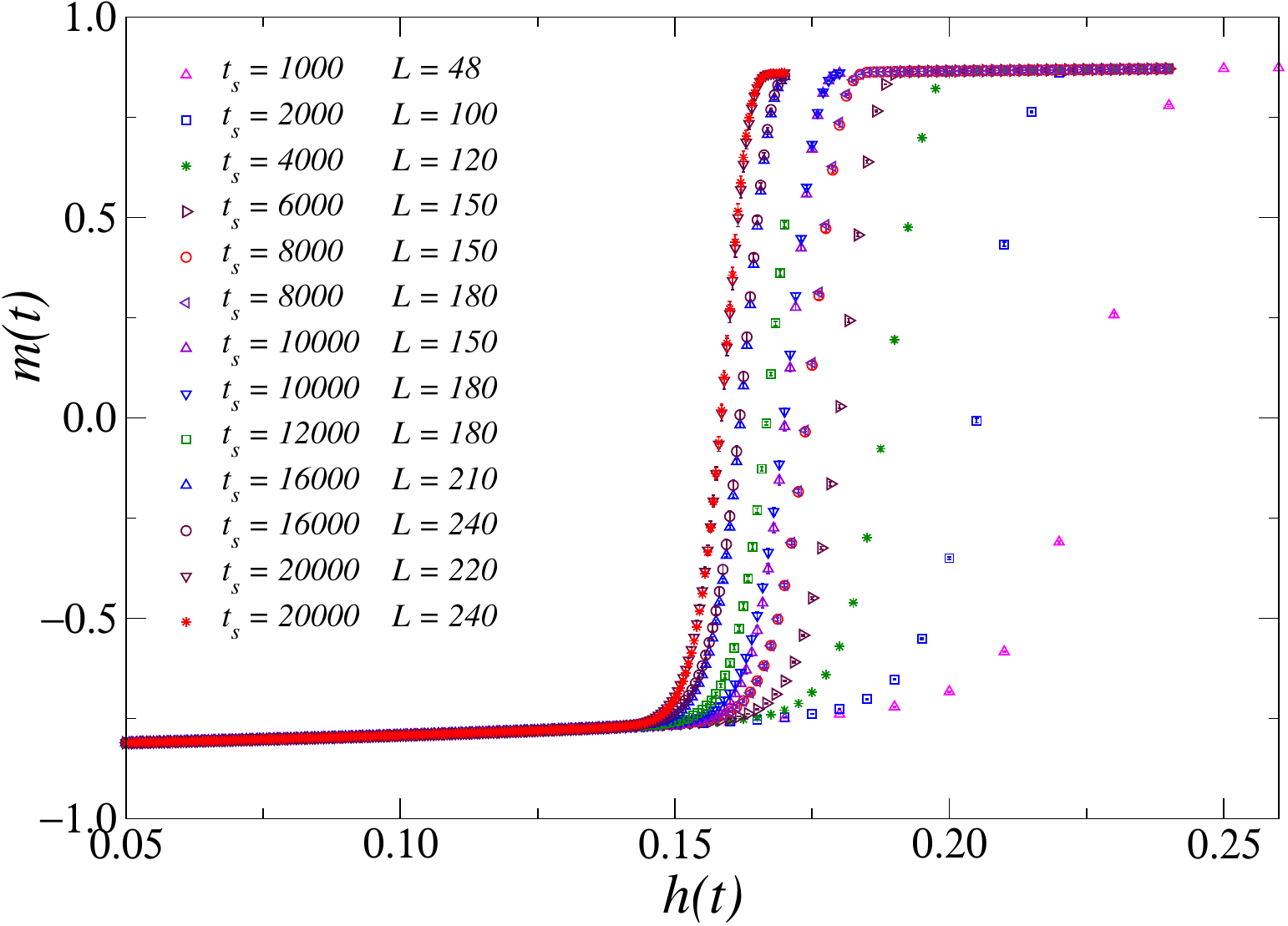}
    \caption{The average magnetization (bottom) and the bond-energy
      density (top) versus $h(t)$ in 3D Ising systems at
      $\beta=1.2\beta_c$, for several values of $t_s$ and $L$.  We
      consider the Metropolis dynamics.  Statistical errors are hardly
      visible on the scale of the figure.  For some values of $t_s$ we
      report data for two lattice sizes, to demonstrate that the
      results provide an accurate approximation of the evolution in
      the infinite-size limit.  }
\label{datad3}
\end{figure}

\section{The 3D thermodynamic limit}
\label{3dTL}

\begin{figure}[tbp]
  \includegraphics[width=0.9\columnwidth, clip]{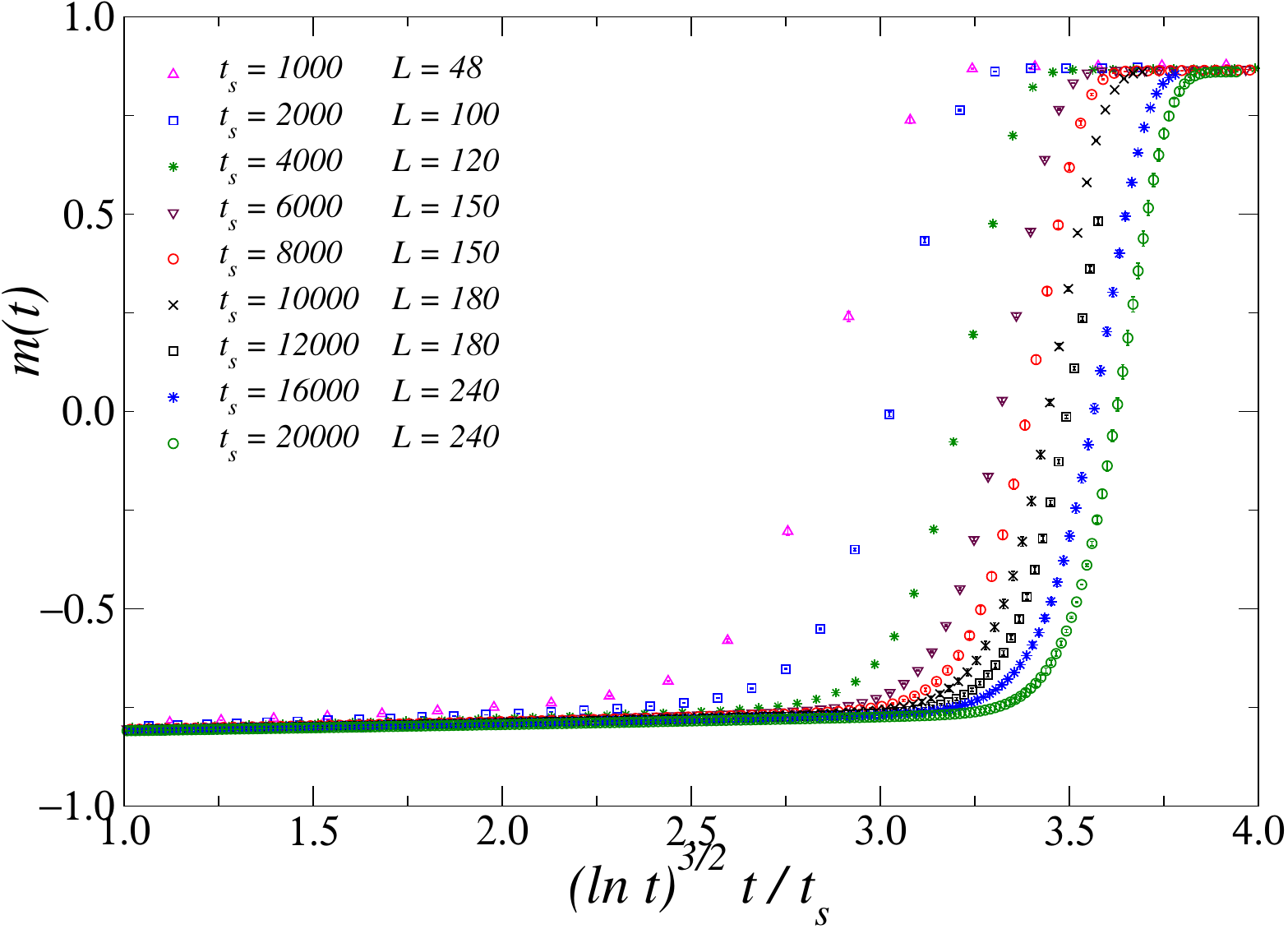}
  \caption{Magnetization $m$ versus $(\ln t)^{3/2} t/t_s$. For each
    $t_s$ we consider the largest value of $L$, so we are essentially
    considering the evolution in the infinite-volume limit.}
\label{firstresc3dtest}
\end{figure}

We now report an analogous study of the KZ dynamics in the TL in three
dimensions. We perform Metropolis KZ simulations at
$\beta=1.2\,\beta_c$ on lattices of size up to $L=240$.  They allow us
to obtain infinite-volume dynamical results for values of $t_s$ up to
$t_s=2\times 10^4$ (the infinite-size limit is observed for $L \gtrsim
1.5 \sqrt{t_s}$, as in the 2D case, see Sec.~\ref{2dTL}).  Results for
the evolution of the magnetization and of the bond-energy density in
the TL are shown in Fig.~\ref{datad3}. For some values of $t_s$ we
report data for two lattice sizes, to demonstrate that the results
provide an accurate approximation of the evolution in the
infinite-size limit.

At first, we have studied if data scale as predicted by the arguments
presented in Sec.~\ref{outsca}, based on the idea that the relevant
mechanism is the nucleation of spherical smooth droplets.  As is
evident from Fig.~\ref{firstresc3dtest}, data do not scale and we
observe a systematic drift of the curves towards larger values of the
scaling variable as $t_s$ increases. Clearly, the arguments of
Sec.~\ref{outsca}, which allowed us to successfully obtain a scaling
picture of the dynamics in the TL in two dimensions, apparently fail
in three dimensions.

\begin{figure}[tbp]
  \includegraphics[width=0.9\columnwidth, clip]{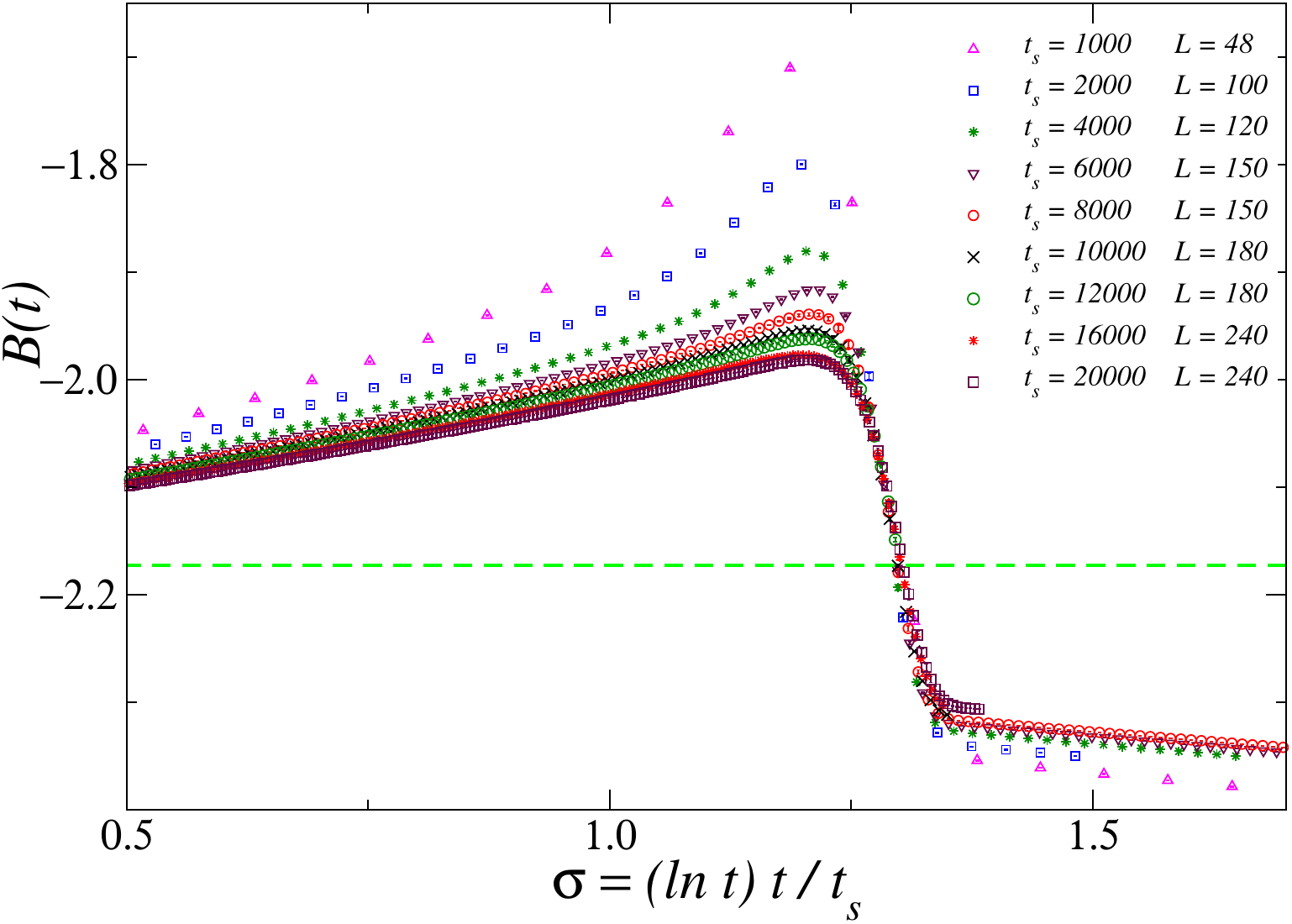}
  \includegraphics[width=0.9\columnwidth, clip]{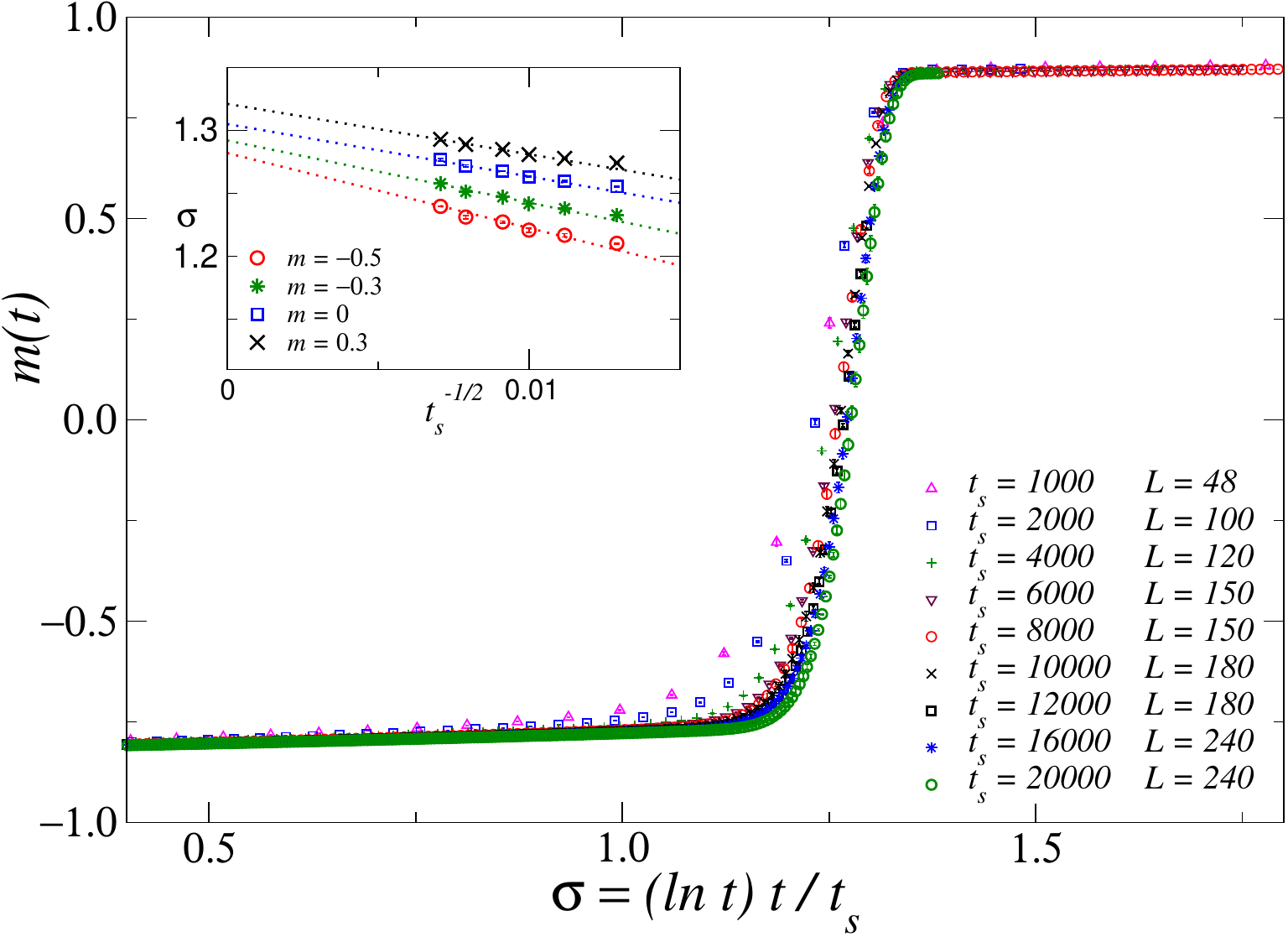}
  \caption{The magnetization $m$ (bottom) and bond-energy density $B$
    (top) for the 3D Ising model, versus $\sigma=t \ln t/t_s$. 
    Results for the Metropolis dynamics at $\beta=1.2\beta_c$.
    The dashed line in the upper panel corresponds to the equilibrium
    energy density at $h=0$. Results approach an asymptotic
    scaling curve with increasing $t_s$. The inset in the lower panel
    shows the values of $\sigma$ where $m(t)$ is equal to 
    the values reported in the legend, as a function of 
    $t_s^{-1/2}$. Data apparently lie on a straight line; the dotted
    lines correspond to linear fits of the data to $a+b\,t_s^{-1/2}$ 
    (in the fits we only include the results with $t_s\gtrsim 8000$).  }
\label{firstresc3d}
\end{figure}

We have therefore studied whether scaling can be observed by changing
the scaling variable, and, in particular, by changing the power of
$\ln t$ in the definition of $\sigma$. As shown in
Fig.~\ref{firstresc3d}, scaling is observed if we consider
\begin{eqnarray}
  \sigma = {t \ln t \over t_s}.
 \label{sdef3d} 
\end{eqnarray}
Indeed, the infinite-volume magnetization appears to scale as
\begin{eqnarray}
M_\infty(t,t_s) \approx {\cal
  M}_\infty(\sigma) + O(t_s^{-\varepsilon}),
    \label{mtsca3d}
\end{eqnarray}
with power-law corrections that are consistent with
$\varepsilon\approx 1/2$, see the inset in the lower panel of
Fig.~\ref{firstresc3d}.  An analogous scaling behavior is also
observed for the bond-energy density $B$, as shown in the upper panel
of Fig.~\ref{firstresc3d}.  The apparent large-$t_s$ collapse of the
data provides a strong support to the scaling variable
(\ref{sdef3d}). However we cannot exclude a scaling behavior in terms
of $t \ln^\kappa t/t_s$ with $\kappa$ sufficiently close to one.  Data
suggest that the accuracy on the value $\kappa = 1$ is approximately
10\%.  Note that the large-$t_s$ scaling functions are smooth with no
singularities, at variance with what occurs in the 2D case. We have at
present no direct interpretation of the observed scaling behavior in
terms of the variable (\ref{sdef3d}). However, if the relevant
mechanism is the nucleation of droplets, we would predict that the
time needed to create a droplet of size $R$ should increase as $\ln^3
R$, which, in turn, would suggest that these 3D droplets are fractal
objects.

Also in three dimensions we can define a critical $h_*$. For instance, 
it might be defined as the magnetic field for which $m(t)=0$. As a
consequence of the definition (\ref{sdef3d}), 
$h_* >0$  decreases logarithmically as
\begin{equation}
  h_* \sim {1\over \ln t_s},
  \label{hstar3D}
  \end{equation}
to be compared with the behavior $h_* \sim 1/(\ln t_s)^2$ obtained in
two dimensions.

\begin{figure}[htbp]
  \includegraphics[width=0.9\columnwidth, clip]{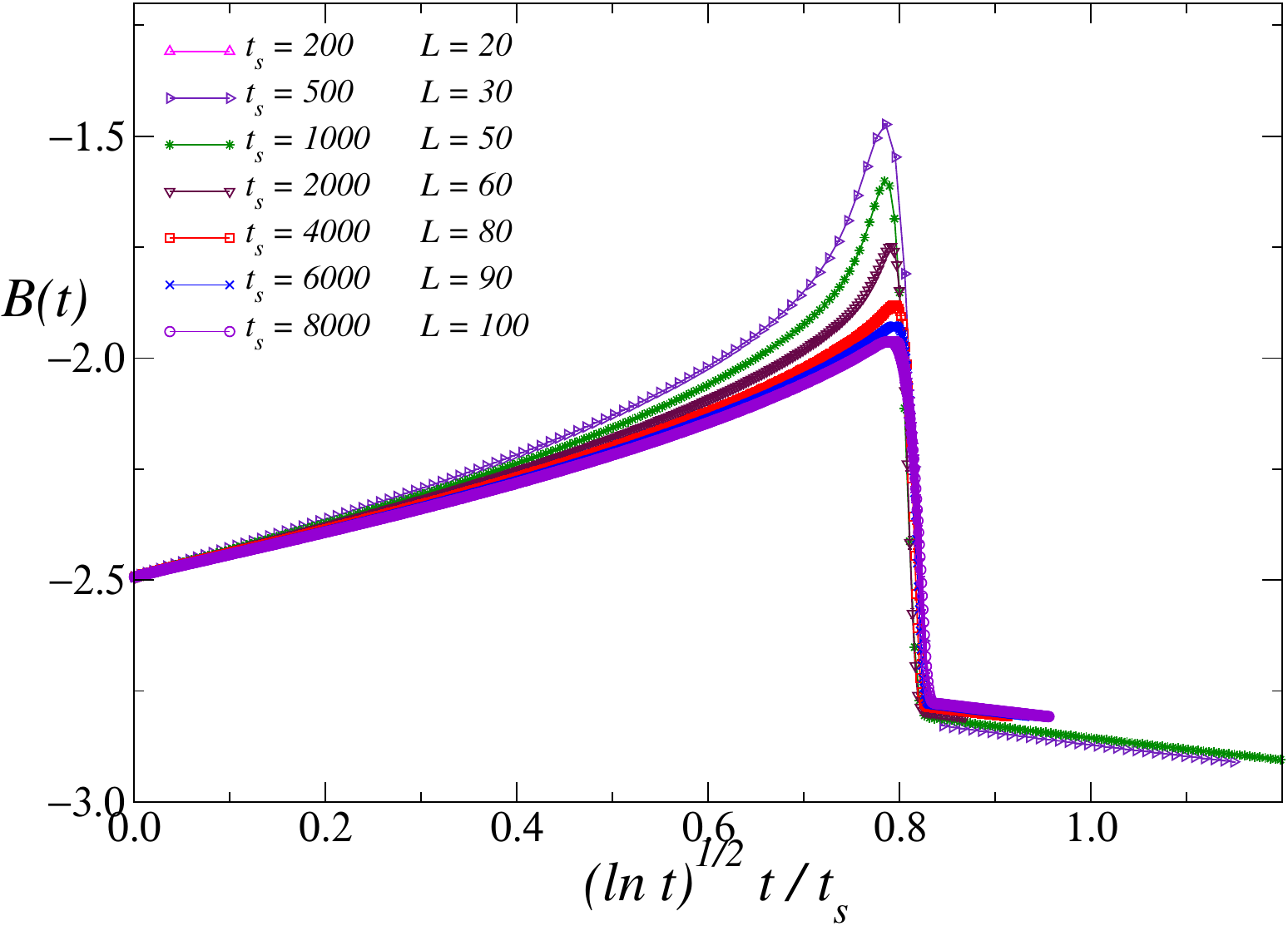}
  \includegraphics[width=0.9\columnwidth, clip]{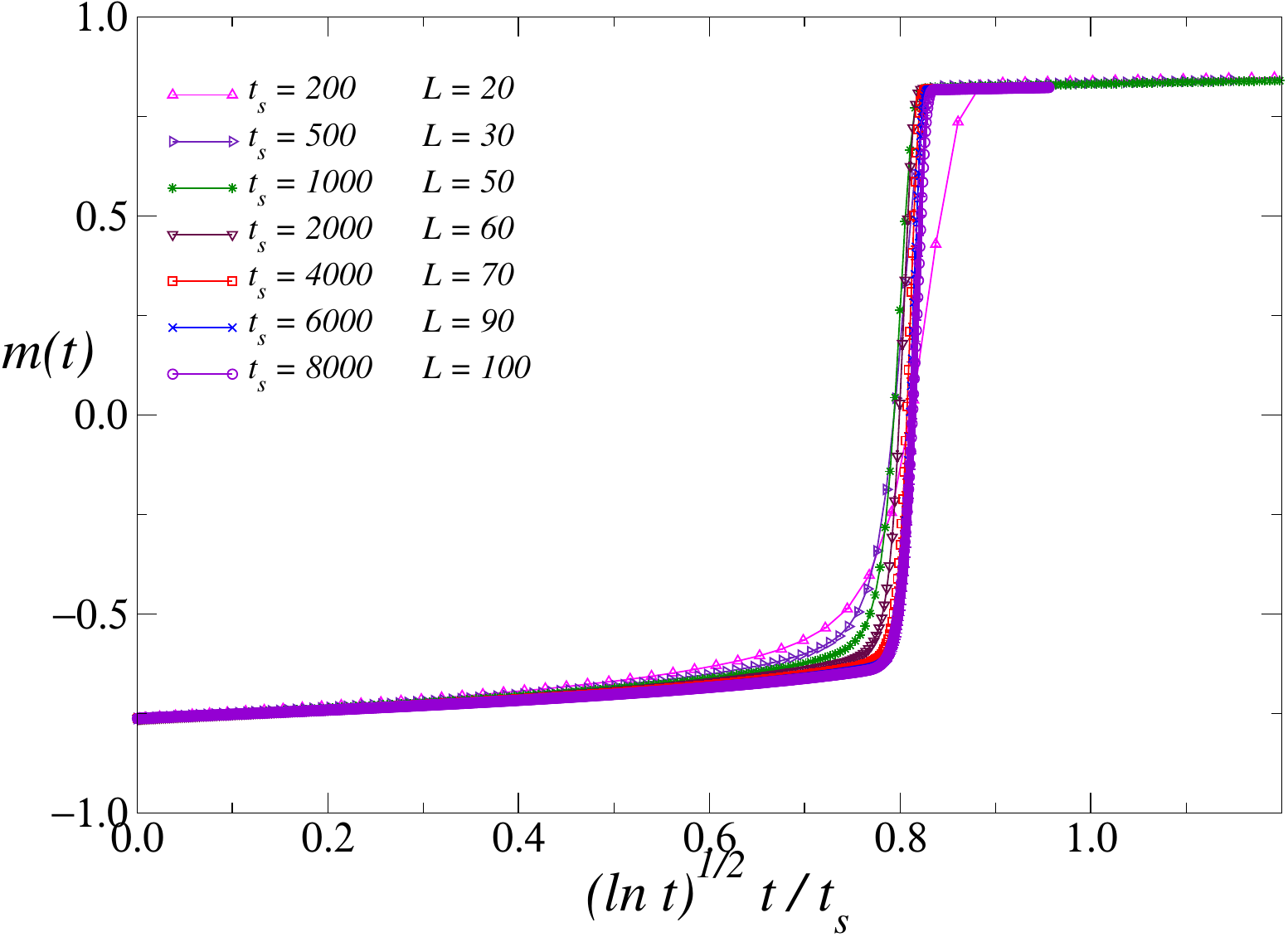}
  \caption{The magnetization (bottom) and bond energy (top)
      across the magnetic first-order transition of the 4D Ising model
      at $\beta=0.18>\beta_c\approx 0.15$, versus $\sigma = t (\ln
      t)^{1/2}/t_s$, up to $t_s=8000$. All data correspond to the
      thermodynamic large-$L$ limit keeping $t_s$ fixed. The data
      appear to approach a unique large-$t_s$ curve, characterized by
      an abrupt change at $\sigma\approx 0.8$.}
\label{4Dtssresc}
\end{figure}

Finally, we conclude this section by briefly reporting some
  results for the 4D Ising model along its magnetic first-order
  transiton line.  We present some results at $\beta=0.18 \approx
  1.2\,\beta_c$
  ($\beta_c=0.149693785(20)$~\cite{LM-23,LXZFD-24}) in the TL.
  Data for the magnetization and the bond energy density up
  to $t_s=8000$ are shown in Fig.~\ref{4Dtssresc}, versus the scaling
  variable $\sigma = h(t) (\ln t)^{\kappa}$. Data suggest $\kappa=1/2$,
  with an accuracy of approximately 20\%.  The observed 
  behaviors are analogous to those observed in three dimensions,
  see Fig.~\ref{firstresc3d}, the only difference being the value 
  of $\kappa$.

\section{Conclusions}
\label{conclu}

We have reported a study of the out-of-equilibrium scaling behavior
occurring in Ising systems when they are slowly driven by an external
time-dependent and spatially homogenous magnetic field $h(t)$
across their magnetic first-order transitions at temperatures below the
temperature $T_c$ where the continuous transition occurs. We 
consider KZ protocols in which the magnetic field $h$ varies as
$h(t)=t/t_s$ with a time scale $t_s$, while the temperature is kept
fixed. We start the dynamics from equilibrium states at a negative
value $h_i<0$, so the magnetization of the system is negative, and
stop the evolution at a positive value of $h(t)$, when the average
magnetization of the system is positive, therefore crossing the
first-order transition at $h=0$. We consider two different purely
relaxational dynamics, the Metropolis and the heat-bath
dynamics. We compute the magnetization and the
bond-energy density as a function of time, and average the results
over a large number of trajectories starting from the equilibrium
ensemble of configurations at $h_i$.

We focus on two different dynamic regimes.  First, we consider the
OFSS limit, defined as the simultaneous $L,\,t,\,t_s\to \infty$ limit 
keeping appropriate scaling variables fixed---they are defined in
Eqs.~(\ref{rudef}) and (\ref{wdef}). In this case the relevant time
scale is the exponentially large time associated with the tunneling
process in which the system changes phase at $h=0$. Then, we study the
behavior in the TL. In this case, we first consider the infinite-size
limit while keeping $t$ and $t_s$, and thus $h(t)$, fixed.  Then, we
identify the scaling behavior in the large-$t_s$ limit.

Our numerical analyses confirm that the time evolution in finite-size
systems with periodic boundary conditions obeys the general OFSS
relations reported in Sec.~\ref{OFSS-theory}.  On the other hand, the
results in the TL show the emergence of a spinodal-like behavior.  The
average magnetization changes sign at a $t_s$-dependent positive
magnetic field $h=h_*>0$. In the large-$t_s$ limit, $h_*$ turns out to
decrease as $1/(\ln t_s)^\kappa$, with $\kappa = 2$ for 2D Ising
systems, $\kappa=1$ for 3D Ising systems, and $\kappa=1/2$ for 4D
Ising systems.  In two dimensions the time evolution of the
magnetization in the large-$t_s$ limit develops a singular behavior as
a function of the scaling variable $\sigma = t (\ln t)^2/t_s$ at a
specific value $\sigma_*$, which is independent of $t_s$ for large
values of $t_s$. Moreover, close to $\sigma_*$ we observe a scaling
behavior in terms of $\hat\sigma = (\sigma - \sigma_*) t_s^\theta$
where $\theta>0$. The exponent $\theta$ turns out to depend on the
temperature and the specific dynamics.  In three dimensions, the
scaling behavior is qualitatively different.  The data appear to scale
as a function of $\sigma = t (\ln t)/t_s$, without developing
asymptotic singularities like those in two dimensions.  An
  analogous behavior is observed in four dimensions, in terms of the
  scaling variable $\sigma = t (\ln t)^{1/2}/t_s$.

We note that the values of $\kappa$ from two to four dimensions may be
interpolated by the simple formula $\kappa=(6-d)/d$, which gives
$\kappa=2,\,1,\,1/2$ for $d=2,\,3,\,4$ respectively.  This may suggest
$\kappa=0$ for $d\ge 6$. Note that $\kappa =0$ is the mean-field
prediction that should hold in the limit $d\to\infty$.  The value
$\kappa = 0$ would be consistent with a standard spinodal picture: In
the KZ evolution in the thermodynamic and large-$t_s$ limits the
system moves across metastable states for sufficiently small values
$h>0$, up to a time corresponding to a finite $h=h_*>0$, when it
suddenly changes phase (see footnote~\ref{footnote1} of
Sec.~\ref{intro}).  Of course, this is a wild speculation that would
require some theoretical support and additional numerical work.
  
The results for the 2D Ising model are analogous to those obtained at
the thermal first-order transition (in this case the temperature is
the driving variable) of the 2D Potts models for a large number of
states, both in the OFSS and TL regimes~\cite{PV-17,PV-24}.  In
particular, the singular behavior of the KZ scaling functions in the
TL observed in 2D Ising systems is analogous to that observed in the
Potts model.

Concerning the results for the 3D Ising model, we note that the
observed out-of-equilibrium TL scaling behavior is not consistent with
the general arguments that assume that the relevant mechanism driving the 
phase change 
is the nucleation of smooth droplets. This
suggests that a different mechanism is at work in three dimensions.
Of course, we stress that the TL scaling that we put forward is only
based on numerical results.  The study of other systems would be very
useful to further corroborate the general scenario we propose, based
on a logarithmic scaling in terms of the driving parameter.  In
particular, it would be useful to verify whether the power of the
logarithm that appears in the definition of the scaling variable
$\sigma$ is the same at other 3D thermal first-order transitions.  It
would be also interesting to investigate more realistic 3D systems, in
which the proposed behavior can be also tested experimentally. For
instance, one could consider fluid systems and KZ protocols driving
them across the liquid-vapor first-order transition, or binary systems
driven across the demixing transition. 

We finally mention that analogous spinodal-like scaling behaviors have
been observed at quantum transitions. Refs~\cite{PRV-25,PRV-25b}
considered the quantum Ising chain in a transverse field $g$ and
considered the KZ unitary dynamics in the presence of a slowly varying
longitudinal magnetic field $h(t)=t/t_s$, driving the system across
the first-order quantum transitions occurring for small values of $g$.
If one defines $h_*$ as the magnetic field, where the longitudinal
magnetization changes sign, one finds a similar logarithmic scaling
law, $h_* \sim (\ln t_s)^{-1}$~\cite{PRV-25,PRV-25b}.

\appendix
\section{Effective model for the dynamics}

A simple effective model for the KZ dynamics in the OFSS
regime can be obtained by generalizing the equilibrium analysis
presented in Ref.~\cite{PV-17b}.  Since on time scales of the order of
$\tau(L)$ the reversal of the sign of the magnetization is essentially
instantaneous, we can consider a simpler coarse-grained
dynamics. First, we assume that the rescaled magnetization
$M(t)={m(t)/m_0}$ takes only two values, $M(t) = \pm 1$. Second, as we
expect the dynamics restricted within each free-energy minimum to be
rapidly mixing, we can assume that the coarse-grained dynamics is
Markovian. Under these conditions, the dynamics is completely
parametrized by the rates $I_+$ and $I_-$ defined by
\begin{eqnarray}
&& P[M(t)= -1 \to M(t + dt) = 1,t]= I_+ dt, \nonumber \\
&& P[M(t)= 1 \to M(t + dt) = -1,t]= I_- dt, \quad
\label{Markov}
\end{eqnarray}
where $P(\cdot,t)$ is the probability of the considered transition at
time $t$. 
For $t = 0$, i.e., at $h=0$, the rates can be
related to the average time $\tau(L)$ for which a reversal of the
magnetization is observed:
\begin{equation}
   I_+ = I_- = {1\over \tau(L)}.
\end{equation}
For $t > 0$ we assume a similar relation 
\begin{equation}
    I_+ = {h_+(\Phi)\over \tau(L)},  \qquad 
    I_- = {h_-(\Phi)\over \tau(L)},
\end{equation}
with $h_\pm(0) = 1$.
We expect the two functions $h_+(\Phi)$ and $h_ (\Phi)$
to be functions of $\Phi$ in the OFSS regime. This assumption has been 
verified for a specific values of $\Phi$ in the 2D case (we fix
$\beta = 1.2\beta_c$ and consider the heat-bath dynamics).
We consider $\Phi = 8.37$ and determine 
the average time $\overline{\tau}$ the system
takes to reverse the sign of the magnetization, when it is prepared 
in a negatively-magnetized state thermalized at $h=0$. We find 
$h_+ = \tau(L)/\overline{\tau} = 37(1)$, 34(2), 40(2) for 
$L=10,12,14$, indicating that $h_+(\Phi)$  only depends on the scaling 
variable $\Phi$.

\begin{figure}[tbp]
\includegraphics[width=0.9\columnwidth, clip]{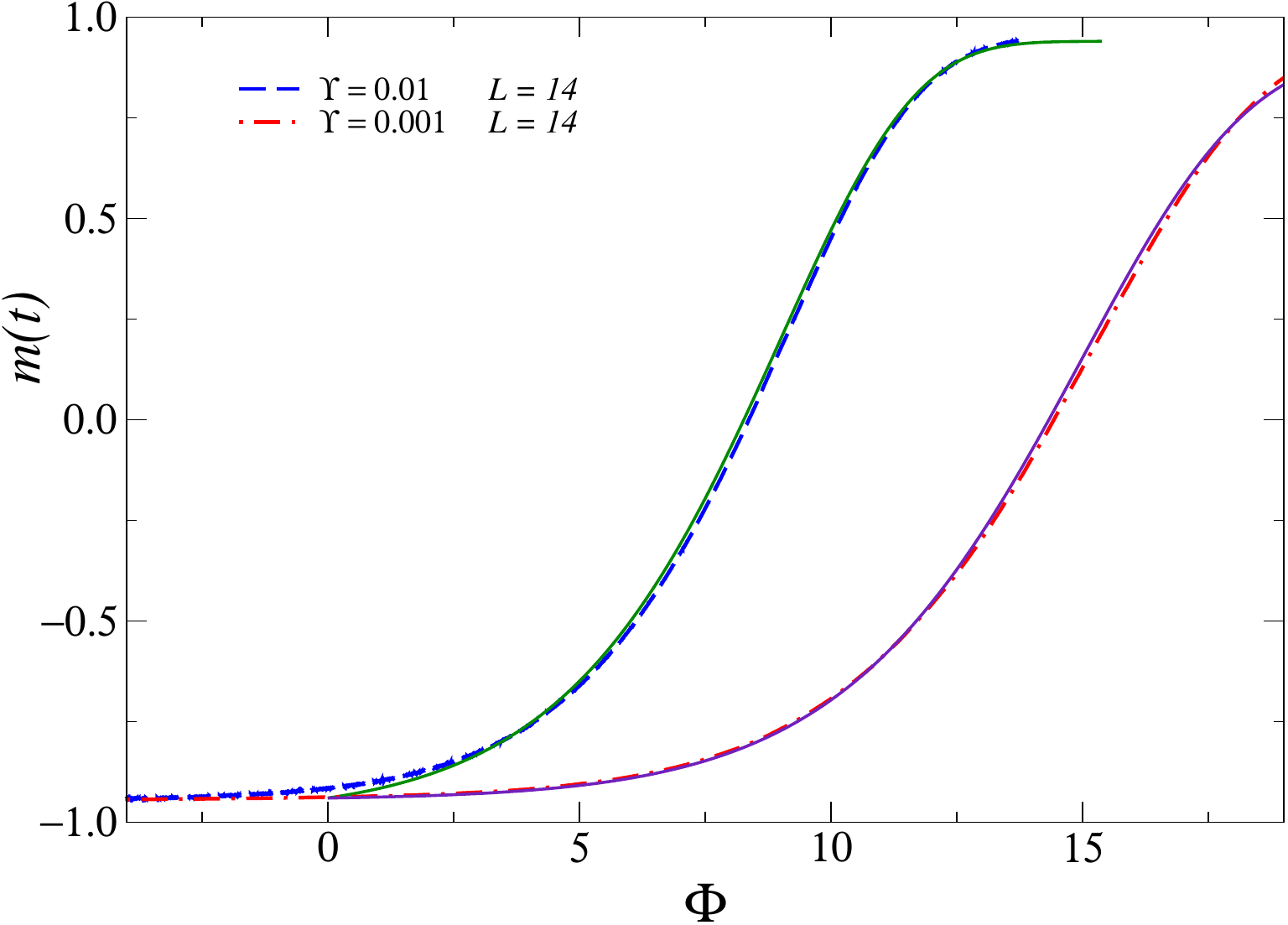}
    \caption{Average time evolution of the magnetization under a
      heat-bath KZ dynamics at $\beta=1.2\beta_c$ (2D Ising model) and
      $L=14$ for $\Upsilon=0.01$ and 0.001.  Results are compared with
      the predictions of the effective model, see
      Eqs.~(\ref{n-prediction}) and (\ref{m-prediction}).  }
\label{App:FSSmagn}
\end{figure}

We now consider the fraction $n(\Phi,\Upsilon)$ of systems that have
positive magnetization at the time corresponding to $\Phi$ for a given
value of $\Upsilon$.  As a consequence of Eq.~(\ref{Markov}), the
function $n(\Phi,\Upsilon)$ satisfies the differential equation
\begin{equation}
{dn\over d\Phi} = \Upsilon \left[(1 - n) h_+(\Phi) + 
     n h_-(\Phi) \right],
\label{effectivemodel}
\end{equation}
with $n(\Phi=0,\Upsilon) = 0$. In the typical range of $\Phi$ in which 
the system magnetization changes sign, we have $h_+(\Phi) \gg h_-(\Phi)$ 
(in our simulations we have never observed transitions from the 
stable positively-magnetized state towards the 
metastable state), so we can neglect $h_-(\Phi)$, obtaining 
\begin{equation}
   n(\Phi,\Upsilon) = 1 - \exp\left[- \Upsilon
    \int_0^\Phi d\Psi\, h_+(\Psi)\right].
\label{n-prediction}
\end{equation}
The magnetization is obtained using 
\begin{equation}
   m(\Phi,\Upsilon) = m_0 [2 n(\Phi,\Upsilon) - 1].
\label{m-prediction}
\end{equation}
To obtain predictions, we should somehow obtain a reasonable approximation
for the function $h_+(\Phi)$. At equilibrium the ratio $I_+/I_-$
is related with the difference of the free energies of the stable 
and metastable phases. If we approximate this difference  with 
$2 m_0 h L^d$, we end up with the approximate formula
\begin{equation}
  {I_-\over I_+} \approx e^{-2\beta m_0 \Phi}.
\label{eq:ratioI}
\end{equation}
If we take into account the symmetry under $h\to -h$,
Eq.~(\ref{eq:ratioI}) suggests the parametrizations
\begin{equation}
   h_+(\Phi) = e^{\beta m_0 \Phi}  g_+(\Phi), \quad
   h_-(\Phi) = e^{-\beta m_0 \Phi} g_-(\Phi), 
\end{equation}
where the functions $g_\pm(\Phi)$ are expected to be slowly varying
functions of $\Phi$.  This is confirmed by the numerical data. We have
determined numerically the function $g_+(\Phi)$ in the 2D model
(heat-bath dynamics) for $L=14$ at $\beta = 1.2\beta_c$. For
$\Phi\lesssim 20$ it can be interpolated as
\begin{equation}
g_+(\Phi) = 1 - 0.076115 \, \Phi + 0.001386\, \Phi^2. 
\end{equation}
We can then use Eq.~(\ref{m-prediction}) to predict the behavior of
the magnetization for $\Upsilon = 0.01$ and 0.001, which are the two
values of the scaling variable that we have considered in
Sec.~\ref{numresofss}.  Results are reported in
Fig.~\ref{App:FSSmagn}. The agreement is excellent, which is not
obvious given that we consider two values of $\Upsilon$ that differ by
a factor of 10. These results confirm that the coarse-grained model
effectively takes into account the main features of the KZ dynamics.

\end{document}